\documentclass[journal=jctcce,manuscript=article]{achemso} 
\usepackage{adjustbox}
\usepackage{amsmath}
\usepackage{amssymb}
\usepackage{bm}  
\usepackage{booktabs}
\usepackage{color}
\usepackage{dcolumn} 
\usepackage{etoolbox}
\usepackage[T1]{fontenc}
\usepackage{framed}
\usepackage{geometry}
\usepackage{graphicx}  
\usepackage{hhline}
\usepackage{hyperref}
\usepackage[utf8]{inputenc}
\usepackage{lipsum}
\usepackage{longtable}
\usepackage[version=3]{mhchem} 
\usepackage{multirow}
\usepackage{rotating}
\usepackage{subfigure,amsmath,verbatim,moreverb}
\usepackage{tabularx}
\usepackage{threeparttable} 
\usepackage{xcolor}

\definecolor{bluegreen}{rgb}{0,0.2,0.8}

\UseRawInputEncoding

\AtBeginEnvironment{align}{\setcounter{subeqn}{0}}
\newcounter{subeqn} %

\newcommand{\R}{\mathbf{r}}

\author{Abhishek Bhattacharjee}
\email{abhishekbhattacharjee179@gmail.com ,abhishek.bhattacharjee@niser.ac.in}
\affiliation{School of Physical Sciences, National Institute of Science Education and Research, An OCC of Homi Bhabha National Institute, Bhubaneswar 752050, India}
\author{Subrata Jana}
\email{subrata.jana@umk.pl, subrata.niser@gmail.com}
\affiliation{Institute of Physics, Faculty of Physics, Astronomy and Informatics, Nicolaus Copernicus University in Toru\'n, ul. Grudzi\k{a}dzka 5, 87-100 Toru\'n, Poland}
\author{Szymon \'Smiga}
\affiliation{Institute of Physics, Faculty of Physics, Astronomy and Informatics, Nicolaus Copernicus University in Toru\'n, ul. Grudzi\k{a}dzka 5, 87-100 Toru\'n, Poland}
\author{Prasanjit Samal}
\affiliation{School of Physical Sciences, National Institute of Science Education and Research, An OCC of Homi Bhabha National Institute, Bhubaneswar 752050, India}

\title[An \textsf{achemso} demo]
  {Nonlocal Orbital-Free Kinetic Energy Functional from the Jellium-with-Gap Model for Finite Systems}

\begin{document}

\begin{tocentry}
\includegraphics[width=8 cm, height=4.7 cm]{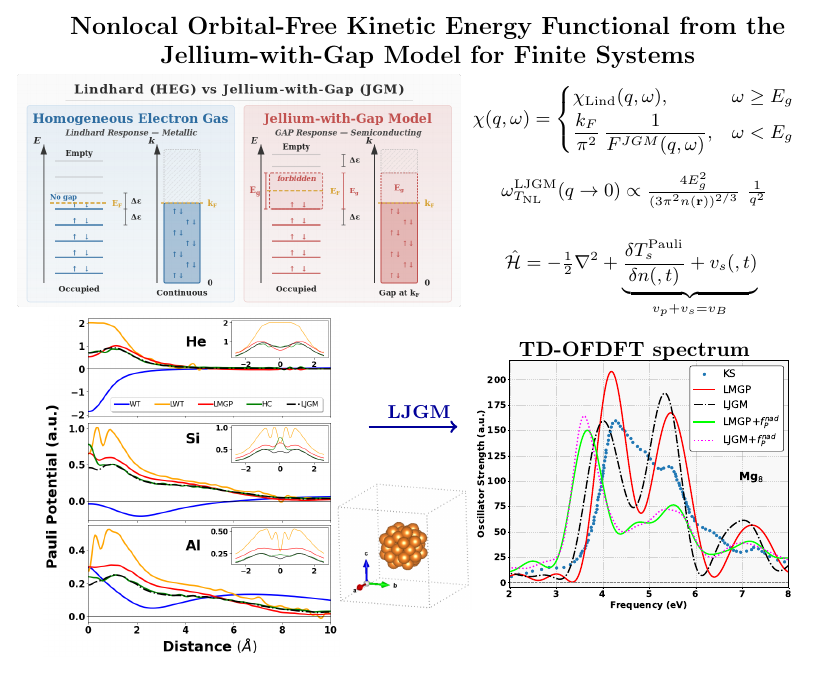}
\end{tocentry}

\begin{abstract}
The quasi-linear scaling of orbital-free density functional theory (OF-DFT) with system size makes it a computationally efficient alternative to conventional Kohn--Sham density functional theory for many condensed-matter applications. However, its applicability remains limited, particularly for finite systems such as molecular clusters, due to the lack of accurate kinetic energy density functionals. In this context, the development of nonlocal kinetic energy density functionals (NL-KEDFs) has significantly advanced the practical utility of OF-DFT. Here, following an alternative formulation based on the linear-response kernel derived from the jellium-with-gap model (JGM), we develop an NL-KEDF capable of accurately describing the diverse density regimes characteristic of finite systems, including molecular clusters. Benchmark calculations, together with an analysis of the corresponding Pauli potentials, demonstrate that the proposed functional achieves higher accuracy than state-of-the-art orbital-free approaches for finite systems. Furthermore, the optical properties computed using the present method show good agreement with reference results, highlighting its reliability. These results indicate that the proposed NL-KEDF provides a robust and efficient framework for extending OF-DFT to finite systems, with potential implications for nanomaterial design and a deeper understanding of nanoscale phenomena. 
\end{abstract}

\section{Introduction \label{sec:Intro}}

Over the past two decades, metal nanoparticles and clusters have garnered significant interest due to their exceptional potential in medicine, sensing, optics, molecular electronics, and catalysis \cite{Cluster-2-D1CP04070G,Cluster-1}. Transition metal nanoclusters, in particular, stand out because they classify as an intermediate state between molecules and bulk solids, defying well-established structural rules~\cite{Cluster-4}. Their size-dependent physical and chemical properties differ markedly from bulk materials, making their study challenging and rewarding~\cite{Cluster-5_PhysRevLett.79.1873,Cluster-6_Au-nano2013}. Understanding their nucleation and growth processes is essential for developing synthetic methods that achieve precise structures and sizes~\cite{Cluster-3-Cortese2018,Cluster-7-Al8_KS_study,Cluster-8-Si-optical-study_2018,Random_structures_OFDFT_Witt2021}.

Electronic structure calculations of such nano-materials are mainly conducted within the Kohn-Sham (KS) density functional theory (DFT) framework~\cite{Cluster-1,Cluster-2-D1CP04070G,Cluster-3-Cortese2018,Cluster-4}. Albeit their sophistication and precise prediction of material properties, the substantial computational cost of KS-DFT triggers other cheap alternatives such as Orbital-Free-DFT (OF-DFT)~\cite{MKH-2024_RevArticle_DFT-2, Rev-article-OFDFT, Della2022Orbital}. 
The OF-DFT has reemerged as a powerful theoretical framework for large-scale simulations, including systems with up to a million atoms~\cite{PROFESS:software,DFTpy:Shao_2020}, warm dense matter~\cite{WDM-1-PhysRevLett.113.155006,WDM-2-PhysRevLett.121.145001,WDM-3-PhysRevLett.111.175002,Large-scale-DFT}, plasmonic\cite{Della2022Orbital, Della2022Advances} and atomic systems~\cite{WGC-Decomposed_Carter_molecoule, YUK1,Yuk2,yuk3}. 

The central aspect of the effectiveness of OF-DFT lies in the accurate approximation of the non-interacting kinetic energy density functional (KEDF), denoted as $T_s[n]$. This functional can be expressed in a generic form as a combination of semilocal and nonlocal contributions as,

\begin{eqnarray}
    T_s[n] &=& \underbrace{T_{\text{TF}}[n] + T_{\text{vW}}[n]}_{\text{semilocal}} 
           +\underbrace{T_{\text{NL}}^{\alpha,\beta}\left[n(\mathbf{r}), n(\mathbf{r}'), \omega_{T_{\text{NL}}}(\mathbf{r}, \mathbf{r}', n(\mathbf{r}))\right]}_{\text{nonlocal}}~,
    \label{eq:intro_eq1}
\end{eqnarray}
where the corresponding kinetic potential is given by~,
\begin{eqnarray}
v_{T_s}(\mathbf{r}) &=& \underbrace{v_{T_{\text{TF}}}(\mathbf{r}) + v_{T_{\text{vW}}}(\mathbf{r})}_{\text{semilocal}} 
+ \underbrace{v_{T_{\text{NL}}}^{\alpha,\beta}(\mathbf{r}, \mathbf{r}')}_{\text{nonlocal}} \;.
\label{intro_eq2}
\end{eqnarray}

Here, the $T_{\mathrm{TF}}[n]$ ($T_{\mathrm{vW}}[n]$) and $v_{T_{\mathrm{TF}}}(\mathbf r) (v_{T_{\mathrm{vW}}}(\mathbf r))$ are the Thomas--Fermi (von Weizs\"acker (vW)) kinetic energy functionals and potentials, respectively~\cite{bookdft_Parr-Yang-1989}. The $T_{\text{NL}}^{\alpha,\beta}$ or $v_{T_{\mathrm{NL}}}^{\alpha,\beta}(\mathbf r)$ is the nonlocal component that depends on a kernel $\omega_{T_{\text{NL}}}(|\mathbf{r} - \mathbf{r}'|, k_F)$. The parameters $\alpha$ and $\beta$ modulate the nonlocal contribution.

Unlike semilocal variants of KEDFs~\cite{VW:vonweizsacker1935zur,JonesAndGunnarsson,Rationalp-(2019)PhysRevB.100.165111,Airy-Gas-KEDF-Lucian-PhysRevB.79.115117,PBEint_bench_Lucian,computation7040065,LKT-SBTrichey-GGA_PhysRevB.98.041111,LKTF-SBTrichey-GGA_PhysRevB.101.075116,PGSL_2018_doi:10.1021/acs.jpclett.8b01926,PGSL_2019_assessment,SG4-PhysRevB.93.045126}, which only exploit the nearsightedness of electronic systems \cite{nearsightedness_Kohn_2005}, the localization of the electron density are generally better described by nonlocal KEDFs (NL-KEDFs)~\cite{review-article-OFDFT-LargeScale,Rev-article-OFDFT}. 
Other than those, recent trends are also focused on the machine-learned KEDFs, which represent a rapidly developing class of functionals that have gained substantial attention in recent years~\cite{ML_KEDF_2022_JCTC,ML_KEDF_JACS_2025_Hamprecht,ML_KEDF_review_2025_Pavanello,ML_KEDF_FUJINAMI2020}. Recent ML-based nonlocal KEDFs, such as the multi-channel CPN KEDF of Sun and Chen ~\cite{ML_KEDF_CPN_2024_Mohan_Chen}, have demonstrated competitive accuracy against state-of-the-art physics-based nonlocal functionals (e.g., HC) for specific classes of systems including Si and III--V semiconductors. Molecular ML-OFDFT models such as STRUCTURES25~\cite{ML_KEDF_JACS_2025_Hamprecht} achieve high accuracy against KS-DFT on broad chemical benchmarks. However, head-to-head comparisons of ML-KEDFs against nonlocal physics-based KEDFs (HC, LMGP, LWT) on general finite systems remain scarce, and as noted in the recent perspective of Pavanello~\cite{ML_KEDF_review_2025_Pavanello}, ML-KEDFs have not yet supplanted KEDFs developed from physical and mathematical reasoning.

Nonlocal KEDFs constructed within the framework of linear-response theory (LRT) of the homogeneous electron gas (HEG)~\cite{WT_1992,WGC_1998,WGC_1999,HC,BLPS-HC,MGP_2018,SM_1993,XMW_PhysRevB.100.205132,YUK1,Yuk2,yuk3,revHC_2021} or based on the jellium-with-gap model (JGM)~\cite{KGAP_2017,KGAP_2018,JGM-10.1063/5.0204957} provide a substantially more refined description of electronic interactions than semilocal approximations.
They significantly improve the accuracy of OF-DFT calculations across a wide range of systems, including bulk solids, clusters, and nanoparticles~\cite{Rev-article-OFDFT,WGC-Decomposed_Carter_molecoule,QiangXu-finite1_2020}.
Such improvements primarily originate from the presence of a nonlocal kernel,
$\omega_{T_{\mathrm{NL}}}(|\mathbf r-\mathbf r'|,k_F)$,
which enables an effective description of spatial density inhomogeneities and nonlocal electronic response effects~\cite{Rev-article-OFDFT,WT_1992}.

Despite these advances, systems characterized by strongly inhomogeneous electron densities, such as molecules and finite clusters-remain particularly challenging for OF-DFT.
For these systems, most density-independent nonlocal KEDFs exhibit large quantitative errors~\cite{LMGP-00}, thereby motivating the development of more general density-dependent nonlocal kernels like as ~\cite{HC,WGC_1999,LMGP-00}.
Although density-dependent formulations offer improved performance, their construction has so far relied predominantly on the Lindhard response function~\cite{bookdft_Dreizler-Gross-1990}, which is strictly valid only for metallic systems.

An important step beyond the Lindhard framework is provided by the JGM kernel~\cite{KGAP_dilectric_origin}, which incorporates a finite electronic gap into the response function.
The JGM kernel satisfies several key exact constraints and exhibits the correct low-$q$ behavior, making it particularly suitable for semiconducting and insulating systems.
To date, the JGM kernel has been successfully employed primarily in the construction of density-independent nonlocal KEDFs~\cite{KGAP_2018,JGM-10.1063/5.0204957}, mainly applicable to bulk systems.

Motivated by the recent development of a density-independent JGM-based functional in Ref.~\cite{JGM-10.1063/5.0204957}, we construct here a density-dependent variant of the JGM kernel tailored for applications to finite atomic and molecular clusters, which constitutes the central objective of the present work.
We systematically investigate both the formal construction and the practical performance of JGM-based density-dependent nonlocal KEDFs, demonstrating their ability to accurately describe localized electron densities while retaining the favorable linear-response characteristics intrinsic to the gap model.

The remainder of this paper is organized as follows.
In Sec.~\ref{sec:Theory}, we describe the theoretical construction of the JGM-based nonlocal KEDF. Section \ref{sec4:Result:pauli_potential} discusses the behaviour of the Pauli potential. In Sec.~\ref{sec3:ResultsDiscussion}, we assess its performance for finite atomic and molecular cluster systems, with particular emphasis on its time-dependent response properties.
Finally, we summarize the main conclusions and discuss future directions in Sec.~\ref{sec5:Conclusion}.

\section{ Theory \label{sec:Theory}}

The present work focuses on the development of the nonlocal KEDFs based on the JGM%
~\cite{KGAP_dilectric_origin,KGAP_2018,JGM-10.1063/5.0204957}
for the description of finite systems, such as atomic and molecular clusters.
To this end, we begin by considering the general form of the density-independent
nonlocal kernel~\cite{WT_1992}~,
\begin{eqnarray}
\omega_{T_{\mathrm{NL}}}(q,n_0)
&=&
\frac{5\,G_{\mathrm{NL}}(\eta)}
{9\,\alpha\,\beta\,n_0^{\alpha+\beta-5/3}},
\label{eq:omega_JGM}
\end{eqnarray}
where $G_{\mathrm{NL}}(\eta)=F^{\mathrm{Lind}}(\eta)-1-3\eta^2$ and
$\eta=q/(2k_{F_0})$.
Here $q$ is the momentum and $k_{F_0}=(3\pi^2 n_0)^{1/3}$ is the Fermi wave vector corresponding to the
average electron density $n_0=\frac{1}{V_{\mathrm{cell}}}\int d\mathbf r\, n(\mathbf r)$. The function $F^{\mathrm{Lind}}(\eta)$ denotes the well-known Lindhard function,
which represents the exact linear-response function of the HEG.
In this formulation, the kernel depends only on the average density $n_0$ and is therefore non-sensitive to spatial variations of the true electron density $n(\R)$.

Density dependence in nonlocal kernels is typically introduced through the
progression $k_{F_0}(n_0)\rightarrow k_F(n(\mathbf r))\rightarrow
\zeta(\mathbf r,\mathbf r')$, where $\zeta$ denotes a two-point Fermi wave vector.
This strategy has been successfully employed in
Refs.~\cite{HC,revHC_2021,WGC_1999} to construct accurate density-dependent
NL-KEDFs.
By replacing the global Fermi wave vector with its local counterpart
$k_F(n(\mathbf r))$, information about system-specific inhomogeneity is embedded
into the response function, resulting in a density-dependent kernel
$\omega_{T_{\mathrm{NL}}}$ of the generalized form
\begin{eqnarray}
\omega_{T_{\mathrm{NL}}}\!\left(q,k_F(n(\mathbf r)),n_0\right)
&=&
\frac{5\,G_{\mathrm{NL}}(\tilde{\eta})}
{9\,\alpha\,\beta\,n_0^{\alpha+\beta-5/3}},
\label{eq:Kernel1}
\end{eqnarray}
with $\tilde{\eta}=q/[2k_F(\mathbf r)]$.
The nonlocal kernel retains the same formal structure as in the density-independent
case,
\begin{equation}
G_{\mathrm{NL}}(\tilde{\eta})
=
F(\tilde{\eta})-1-3\tilde{\eta}^2.
\label{eq:GNL_general}
\end{equation}

Upon adopting the JGM response function, Eq.~\eqref{eq:GNL_general} leads to the
local jellium-with-gap model (LJGM) kernel,
\begin{equation}
G_{\mathrm{NL}}^{\mathrm{LJGM}}(\tilde{\eta},\Delta(\mathbf r))
=
F^{\mathrm{LJGM}}(\tilde{\eta},\Delta(\mathbf r))
-1-3\tilde{\eta}^2,
\label{eq:GNL_LJGM}
\end{equation}
where the function $F^{\mathrm{LJGM}}$ is given by
%
\begin{eqnarray}
\frac{1}{F^{\mathrm{LJGM}}(\tilde{\eta},\Delta)}
&=&
\frac{1}{2}
-
\frac{\Delta}{8\tilde{\eta}}
\left[
\tan^{-1}\!\left(\frac{4\tilde{\eta}+4\tilde{\eta}^2}{\Delta}\right)
+
\tan^{-1}\!\left(\frac{4\tilde{\eta}-4\tilde{\eta}^2}{\Delta}\right)
\right]
\nonumber \\
&&
+
\left(
\frac{\Delta^2}{128\tilde{\eta}^3}
+
\frac{1}{8\tilde{\eta}}
-
\frac{\tilde{\eta}}{8}
\right)
\ln\!\left[
\frac{\Delta^2+(4\tilde{\eta}+4\tilde{\eta}^2)^2}
{\Delta^2+(4\tilde{\eta}-4\tilde{\eta}^2)^2}
\right],
\label{eq:LJGM_kernel}
\end{eqnarray}
%
with the local gap parameter defined as $\Delta(E_g;\mathbf r)=2E_g/k_F^2(\mathbf r)$.

The corresponding LJGM nonlocal kinetic potential is given by
\begin{equation}
v_{T_{\mathrm{NL}}}^{\mathrm{LJGM}}(\mathbf r)
=
\frac{1}{n^{1/6}(\mathbf r)}
\mathcal{F}^{-1}
\!\left[
\mathcal{F}\!\left[n^{5/6}(\mathbf r)\right]
\omega_{T_{\mathrm{NL}}}^{\mathrm{LJGM}}(\tilde{\eta},\Delta(E_g;\mathbf r)
\right],
\label{eq:jgm_potential}
\end{equation}
where $\mathcal{F}$ and $\mathcal{F}^{-1}$ denote Fourier and inverse Fourier
transforms, respectively.

A salient feature of the LJGM functional is that its nonlocal potential is derived
from a line-integral formulation along a scaled-density path similar to LMGP~\cite{MGP_2018}, 
while exactly recovering the linear-response behavior of both metallic and semiconducting
systems~\cite{JGM-10.1063/5.0204957}.
The distinctive low-$q$ behavior of the LJGM kernel arises from the fact that
(i) the low-$q$ correction is explicitly density dependent through
$\Delta(\mathbf r)$ and
(ii) the asymptotic form governs the long-range decay of the potential,
\begin{equation}
\omega_{T_{\mathrm{NL}}}^{\mathrm{LJGM}}(q\rightarrow 0)
\propto
\frac{\Delta^2(E_g;\mathbf r)}{\tilde{\eta}^2}
=
\frac{4E_g^2}{(3\pi^2 n(\mathbf r))^{2/3}}
\frac{1}{q^2}.
\label{eq:omega_low_q}
\end{equation}

As a result, the LJGM kernel naturally enforces the correct long-range constraint
required for localized systems, a feature that is absent in most existing KEDFs. The validity of the JGM response as an approximation to the true KS response in semiconducting systems has been independently verified by Moldabekov \emph{et al.}~\cite{Zhandos_Shao_pavanello_Electronic_Structure_2025}, who extracted the static KE kernel $K_s(q)$ directly from KS-DFT calculations on Si and showed that the UEG-with-gap model reproduces the extracted kernel accurately throughout the physically relevant wavenumber range $q \lesssim 2\pi/(2r_{\rm cut})$. This provides a direct, KS-DFT-based justification for adopting the JGM response as the starting point of our LJGM construction.
While other approaches, such as the MGP~\cite{MGP_2018} and LMGP~\cite{LMGP-00}
functionals, introduce similar behavior via empirical modeling terms (See Eq.5 of ref.~\cite{LMGP-00}), the LJGM kernel achieves this through a physically motivated gap-dependent response. 
in fact there clear difference in LMGP and LJGM construction can be seen from the following: 
\begin{equation}
\omega_{T_{\mathrm{NL}}}(q\rightarrow 0) = \frac{1}{q^2}
\begin{cases}
\frac{4\pi A}{N_e^{2/3}}~\text{erf}^2(q)\text{exp}(-aq^2),~~~  & \text{LMGP} (\text{Eq. 5 of ref.}~\cite{LMGP-00}) \\[6pt]
\frac{4E_g^2}{(3\pi n(\R))^{2/3}}, & \text{LJGM}~. \\[6pt]
\end{cases} \nonumber
\end{equation}
Where $\text{erf}^2(q)\text{exp}(-aq^2)$ are modeling function and A is set as 0.2 empirically and $N_e$ is total number of electrons~\cite{LMGP-00}.

However, for practical applications of the LJGM NL-KEDF to finite clusters, a physically well-motivated construction of the gap parameter $E_g$ is required. In contrast to extended systems, where $E_g$ can be related to bulk electronic properties, its definition for finite systems is nontrivial. In Ref.~\cite{JGM-10.1063/5.0204957}, $E_g$ was determined from a semilocal gap model~\cite{Local-Eg}; however, such a construction is not directly transferable to finite clusters.

In the present work, we instead determine $E_g$ by imposing physically grounded criteria:
\begin{enumerate}
    \item[(i)] For one-electron systems and closed-shell two-electron systems where both electrons occupy the same spatial orbital (e.g., H and closed-shell He ), the exact Pauli potential vanishes identically. Although approximate functionals cannot reproduce this condition exactly, we require that the deviation be minimized according to
    \begin{align}
        \min_{E_g} \left\| 
        \Gamma_\theta^{\mathrm{exact}}
        - 
        \Gamma_\theta^{\mathrm{approx}}[E_g]
        \right\|,
        \label{miminization_condition}
    \end{align}
    where $\Gamma_\theta$ denotes either the Pauli kinetic energy density or the Pauli potential for the corresponding system.

    \item[(ii)] The approximation must preserve the Pauli positivity constraint.

    \item[(iii)] The approximation should minimize the density error over a representative benchmark test set.
\end{enumerate}

All three criteria are simultaneously satisfied within the range 
$0.05 \lesssim E_g \lesssim 0.1$. 
Within this interval, we further refine the choice of $E_g$ by minimizing the combined errors in both the electron density and the total energy. A detailed analytical and numerical assessment of these conditions is presented in the following section. 

We emphasize that the value of $E_g$ used in this work is not associated with the physical band gap of any specific cluster ~\cite{KGAP_dilectric_origin}, but as an effective parameter that allows the LJGM kernel to satisfy the Pauli constraints - which are otherwise violated by most existing KEDFs in finite systems. The Pauli positivity condition and the vanishing of the Pauli potential for one-electron systems together restrict $E_g$ to a narrow physical window $0.05 \lesssim E_g \lesssim 0.1$, before any cluster benchmark is invoked. Within this window, we adopt a single universal value $E_g = 0.05$~eV discussed in Sec.~\ref{Eg_min}. The same value also enters the low-q response correction $\omega_{T_{NL}}(q\to0)\propto \Delta^2/q^2 $ where the system-specific behavior is carried predominantly by the local density through $k_F(r)$ and $\Delta(r)$ while $E_g$ fixes only the overall scale. 
\\ 
However, a system-wise tuning of $E_g$ or $\Delta$ would definitely yield further improvements; we deliberately retain a universal value to preserve the parameter-free character of LJGM. 

\section{Behavior of the Pauli Potentials}
\label{sec4:Result:pauli_potential}

The Pauli potential is a central quantity in OF-DFT, as it accounts solely for the effects of the Pauli exclusion principle (PEP) on the kinetic energy. The quality of the electron density in OF-DFT is strongly linked to the behavior of the Pauli potential~\cite{Exact-pauli-Levy_PhysRevA.38.625,MKH-2023_RevArticle_DFT-1,Lucian_Pauli_JCP2025_pot_10.1063/5.0278570}, making it a valuable diagnostic for assessing the accuracy of KEDFs.

Due to its inherently nonlocal character, the exact expression for the Pauli potential requires knowledge of the KS orbitals and eigenvalues~\cite{Pauli_pot}. In OF-DFT, this is circumvented by using approximate expressions that depend only on the electron density. Typically, the Pauli potential is defined as~\cite{Exact-pauli-Levy_PhysRevA.38.625,LKT-SBTrichey-GGA_PhysRevB.98.041111}:
\begin{align}
    v_{T_\theta}(\mathbf{r}) = \frac{\delta T_\theta[n]}{\delta n(\mathbf{r})} = \frac{\delta}{\delta n(\mathbf{r})} \left[ T_s[n] - T_{vW}[n] \right],
\end{align}
where $T_\theta$ is the Pauli kinetic energy, $T_s$ is the non-interacting kinetic energy, and $T_{vW}$ is the von Weizsacker kinetic energy. The exact Pauli potential and its corresponding energy satisfy several known physical constraints for $\quad \forall~\mathbf{r}$~\cite{Exact-pauli-Levy_PhysRevA.38.625,scaling1,scaling2,scaling3,OFDFT-PauliPositive,LKT-SBTrichey-GGA_PhysRevB.98.041111}:

\begin{enumerate}
    \item[(i)] $F_\theta \ge 0$ , \quad $T_\theta \ge 0$ ~~~ {\textit{(Pauli positivity)}};
    
    \item[(ii)] $ v_{T_\theta}, ~~ t_\theta= 0$~~~ \textit{(for single-orbital systems, e.g., H and closed-shell He)}
    
    \item[(iii)] $v_{T_\theta}[n](\mathbf{r}) \ge 0$, \quad $t_\theta[n](\mathbf{r}) \ge 0$,

    \item[(iv)] $v_{T_\theta}[n](\mathbf{r}) \ge \dfrac{t_\theta[n](\mathbf{r})}{n(\mathbf{r})}$ ;
    \item[(v)] $T_{\theta}[n_{\lambda}] = \lambda^2 T_{\theta}[n], \quad v_{T_\theta}[n_\lambda](\mathbf{r}) = \lambda^2 v_{T_\theta}[n](\lambda \mathbf{r}) $~. 

\end{enumerate}
Here, $F_\theta (= F_s - F_{vW}$) is the Pauli KE functional or enhancement factor, $t_\theta (= t_s - t_{vW}$) is the Pauli KE density, and $\lambda \in [0,1]$ is the uniform coordinate scaling factor, which probes functional behaviour in different density regimes.

Proving all the aforementioned Pauli constraints analytically for  NL-KEDFs is not trivial. Instead, we show that LJGM satisfies all relevant Pauli constraints through a combination of asymptotic analysis and numerical verification.
For nonlocal KEDFs, it has been shown that the WT-class functionals violate the Pauli positivity condition (i.e. condition (i)) in the low-density i.e., $n(\mathbf{r}) \to 0$ limit~\cite{Shao(2024)-effective-WT-PhysRevB.110.085129}. In contrast, newer functionals based on gap-model kernels, such as KGAP~\cite{KGAP_2018}, JGM~\cite{JGM-10.1063/5.0204957}, and the proposed LJGM, respect these constraints in all regions. We provide analytical evidence in Appendix~\ref{App1:PauliPositive} that the JGM kernel satisfies Pauli positivity in various asymptotic limits.

\begin{figure}
    \centering
    \includegraphics[scale=0.3]{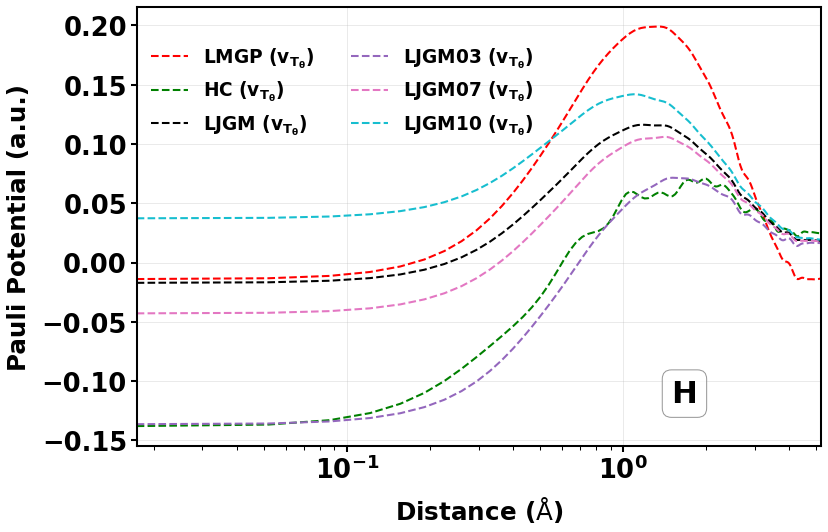}
    \includegraphics[scale=0.3]{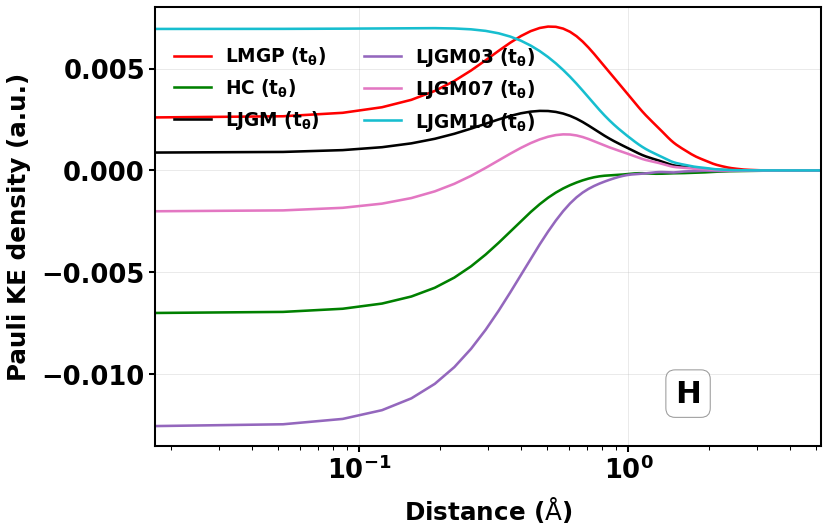}
    \caption{Minimization of Eq.~\ref{miminization_condition} for LJGM $v_\theta$ (upper panel) and $t_\theta$ (lower panel) for H atom (one electron system) through $E_g$. Here, LJGM corresponds to $E_g=0.05$ eV, LJGM03 corresponds to $E_g=0.03$ eV, LJGM07 corresponds to $E_g=0.07$ eV, and LJGM10 corresponds to $E_g=0.1$ eV. 
    }
    \label{fig:placeholder}
\end{figure}

Since the constraint (ii) from the approximate functional, is difficult to satisfy, we apply a minimization procedure, which is shown in Fig.~\ref{fig:placeholder}. We minimize $E_g$ satisfying the condition $\min_{E_g} \left\| \Gamma_\theta^{\mathrm{exact}}
- \Gamma_\theta^{\mathrm{approx}}[E_g]\right\| (\Gamma_\theta \in t_\theta, v_\theta)$ for H atom (one electron system). These results suggest a range 
$0.05 \lesssim E_g \lesssim 0.1$ is necessary to satisfy the constraint (ii). Nevertheless, for practical applications, a fixed value of $E_g$ is useful, which is determined later in this paper.

\begin{figure}
    \centering
    \includegraphics[scale=0.37]{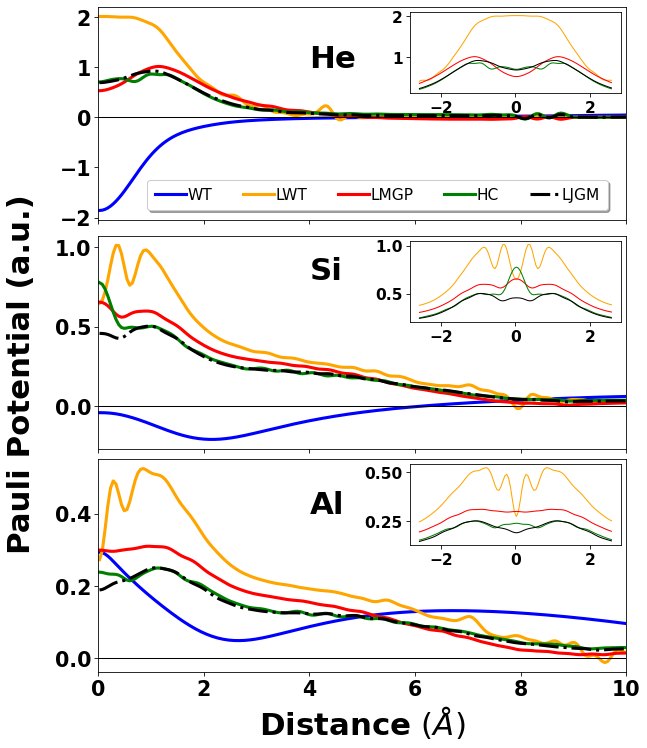}
    \caption{This is a plot of numerical Pauli potentials for the He atom (upper panel),  Si atom (middle panel), and Al atom (lower panel) for various class of functionals highlighting the gradual improvement with higher rungs -- semilocal (TFvW), density-independent NL KEDFs (WT) violating constraint (iii), density-dependent NL-KEDF with Lindhard response (LWT), density-dependent NL-KEDF with correct low-q response (LJGM, HC, LMGP).  
    }
    \label{fig:pauli-plot}
\end{figure}

Further, Fig.~\ref{fig:pauli-plot} 
presents numerical evidences of Pauli positivity for potentials applied to He, Si, and Al atoms with $E_g=0.05$ eV. The value of $E_g$ is justified later in this paper. The results from Fig.~\ref{fig:pauli-plot} indicate that, for He and Si, the WT functional violates the Pauli positivity condition throughout. However, the localized WT (LWT) functional improves upon this and restores positivity, although it exhibits mild oscillations in the density tail region. For the He atom, where the exact Pauli potential is known to vanish, all approximate functionals yield non-zero values, highlighting the sensitivity of light atoms to functional approximations.

\begin{figure}
    \centering
    \includegraphics[scale=0.27]{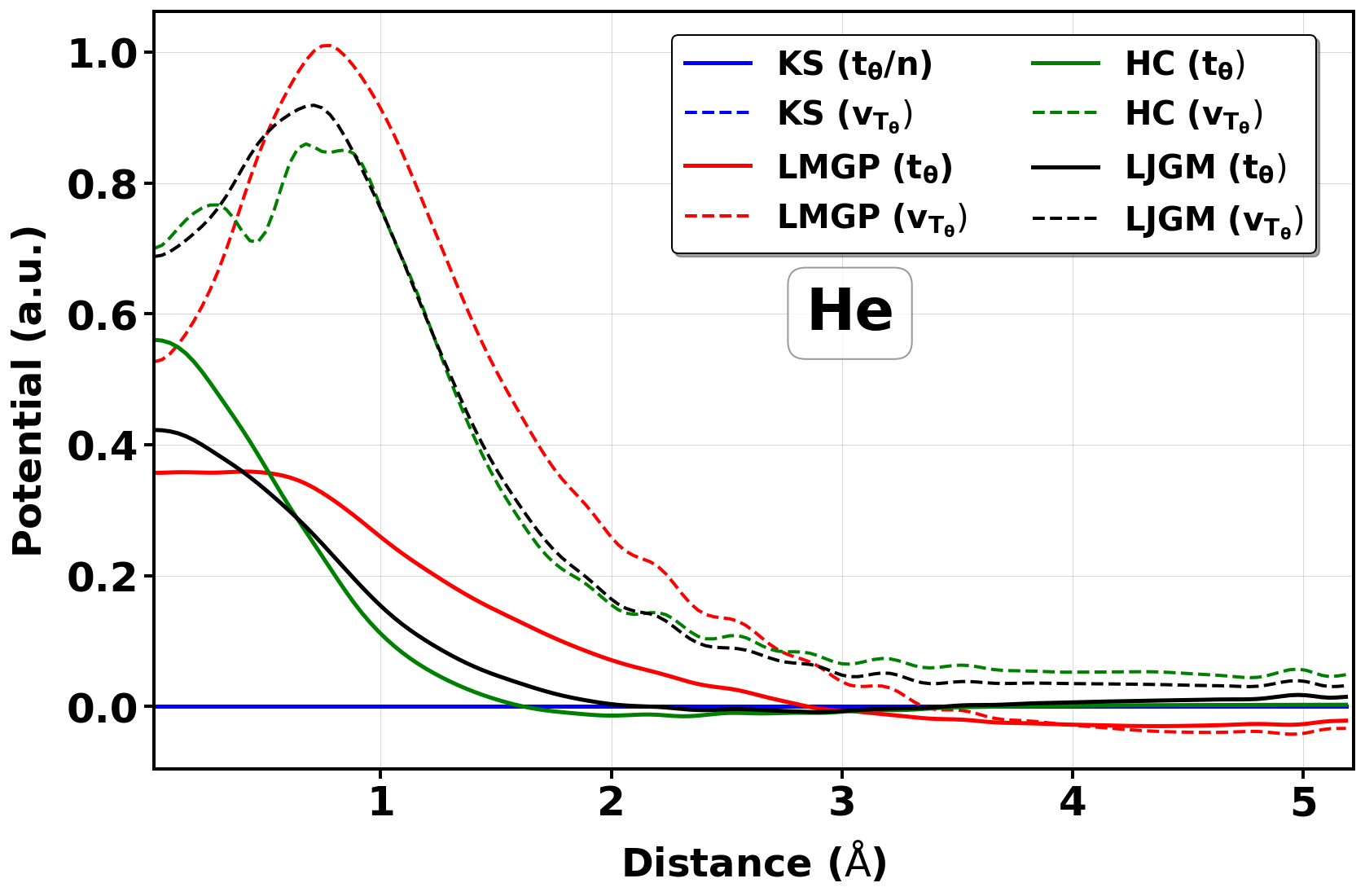}
    \includegraphics[scale=0.27]{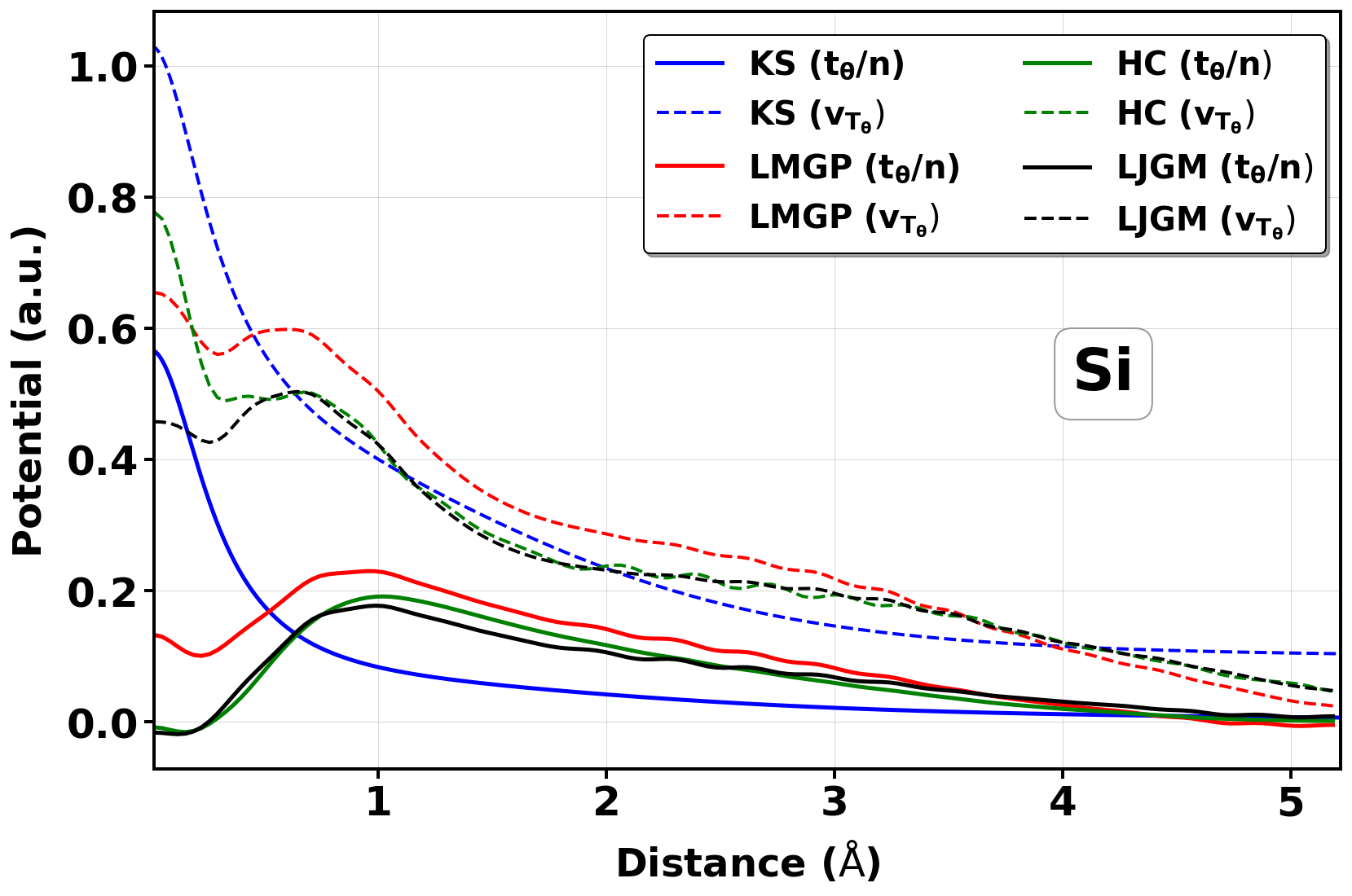}  
    \includegraphics[scale=0.27]{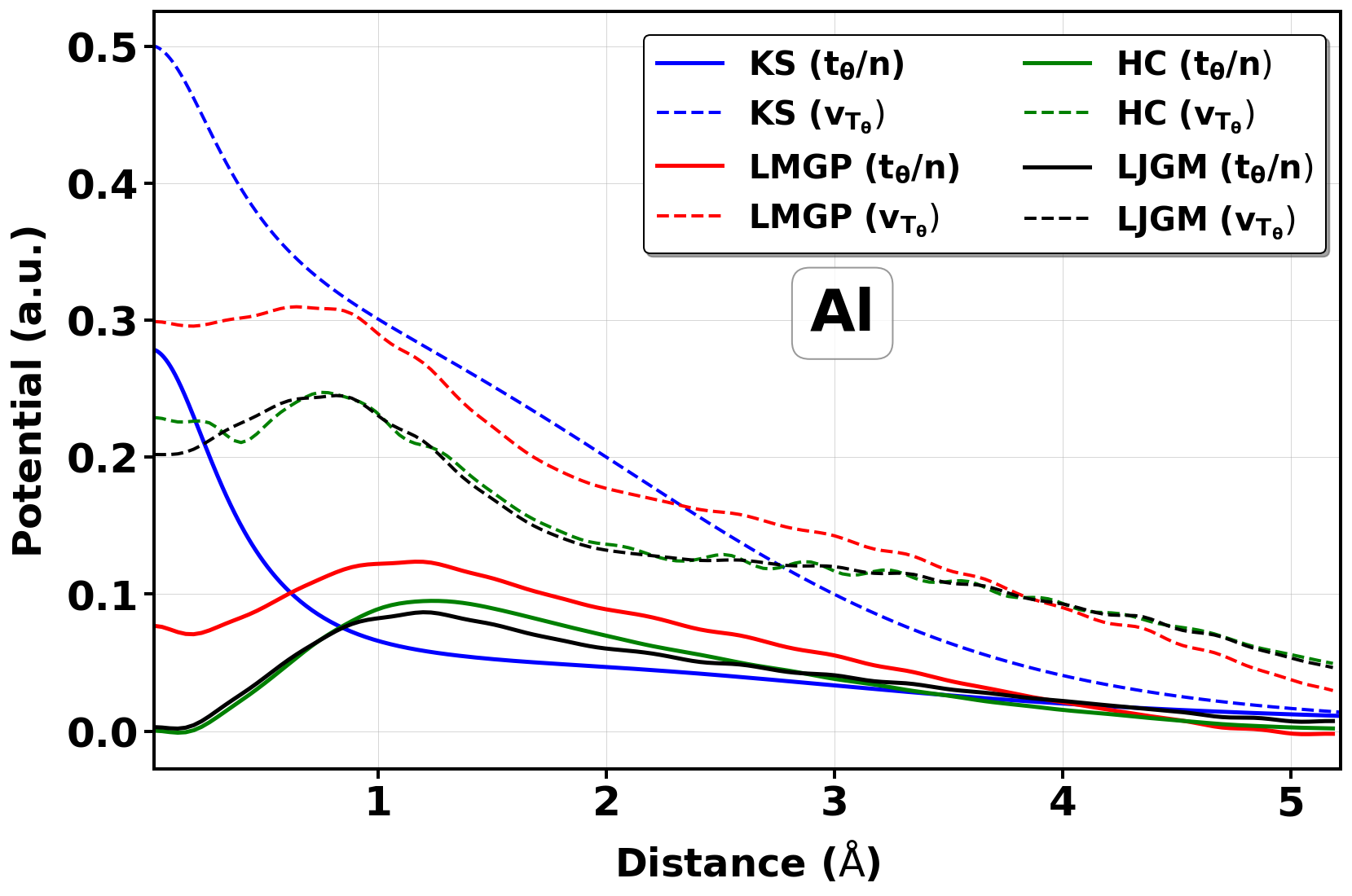}
    \caption{Numerical plot of Pauli potential: $v_{\theta}(\R)$ and $t_{\theta}(\R)/n(\R)$.  The third condition of Pauli constraints is satisfied by LJGM and all other contemporary NL-KEDFs with $E_g=0.05$ eV. 
}
    \label{fig:Tp-LJGM-He}
\end{figure}

The justifications of conditions (iii), together with (iv) are provided in Section~II of Ref.\cite{Exact-pauli-Levy_PhysRevA.38.625} for orbital-based Pauli potential. It serves as an important constraint that needs to be satisfied in general. However, we have proven this condition numerically in Fig.~\ref{fig:pauli-plot} and Fig.~\ref{fig:Tp-LJGM-He}. We have found LJGM and other contemporary NL-KEDFs to satisfy this constraint alike.

The last conditions, i.e., (v) are the scaling constraints which are also satisfied by NL-KEDFs in general. Under coordinate scaling $\R\to \textbf{z}=\lambda\R$, the density scales as $n_{\lambda}=\lambda^3 n(\textbf{z})$. The nonlocal part of kinetic energy (eq.~\ref{eq:intro_eq1}) under coordinate scaling,
\begin{align}
    {T_{NL,}}_\lambda = \int \frac{1}{\lambda^{6}} d^3\textbf{z}d^3\textbf{z}' & \lambda^{3(\alpha+\beta)} n_{\lambda}^{\alpha}~~\lambda^{3} \omega_{\lambda}(k_{F_\lambda}|\lambda\R-\lambda\R'|) n_{\lambda}^{\beta} \nonumber \\
    T_{NL}[n(\R),n(\R')] &= \lambda^2 T_{NL}[n(\textbf{z}),n(\textbf{z}')]
\end{align}
where $\omega\equiv\omega_{T_{NL}}$ for brevity. In the last step we have used $\alpha+\beta=5/3$, as it is the case for LWT, LMGP, and LJGM functionals. Thus we arrive, $T_{NL}[n(\R),n(\R')]= \lambda^2 T_{NL}[n(\textbf{z}),n(\textbf{z}')]$. Similarly, for JGM potential eq.~\ref{eq:jgm_potential} applying the scaling accordingly results,
\begin{align}
    v_{T_{\text{NL}}}(\mathbf{r}) &= \frac{\lambda^{-1/2}~}{n_{\lambda}(\textbf{z})} \mathcal{F_{\lambda}}^{-1} \Big[ \mathcal{F_{\lambda}} \left[ \lambda^{5/2}~n(\mathbf{z})^{5/6} \right] \omega_{\lambda}(\eta_\lambda) \Big]~,
\end{align}
which gives $v_{T_\theta}[n_\lambda](\R) = \lambda^2 v_{T_\theta}[n](\lambda \mathbf{r})$. Thus, in general, the NL-KEDFs obey the scaling constraints when $\alpha+\beta=5/3$. For detailed steps of this calculation, please refer to Appendix~\ref{App2:Scaling-constraint}. 

Note that all these calculations are performed using the pseudopotential code. Technical details of the calculation procedures, software, and pseudopotentials used in this part of the calculations are provided in section~\ref{technical_details}

\section{Application to Real Systems}
\label{sec3:ResultsDiscussion} 

\subsection{Benchmark test set}

\begin{figure}
    \centering
    \includegraphics[scale=0.20]{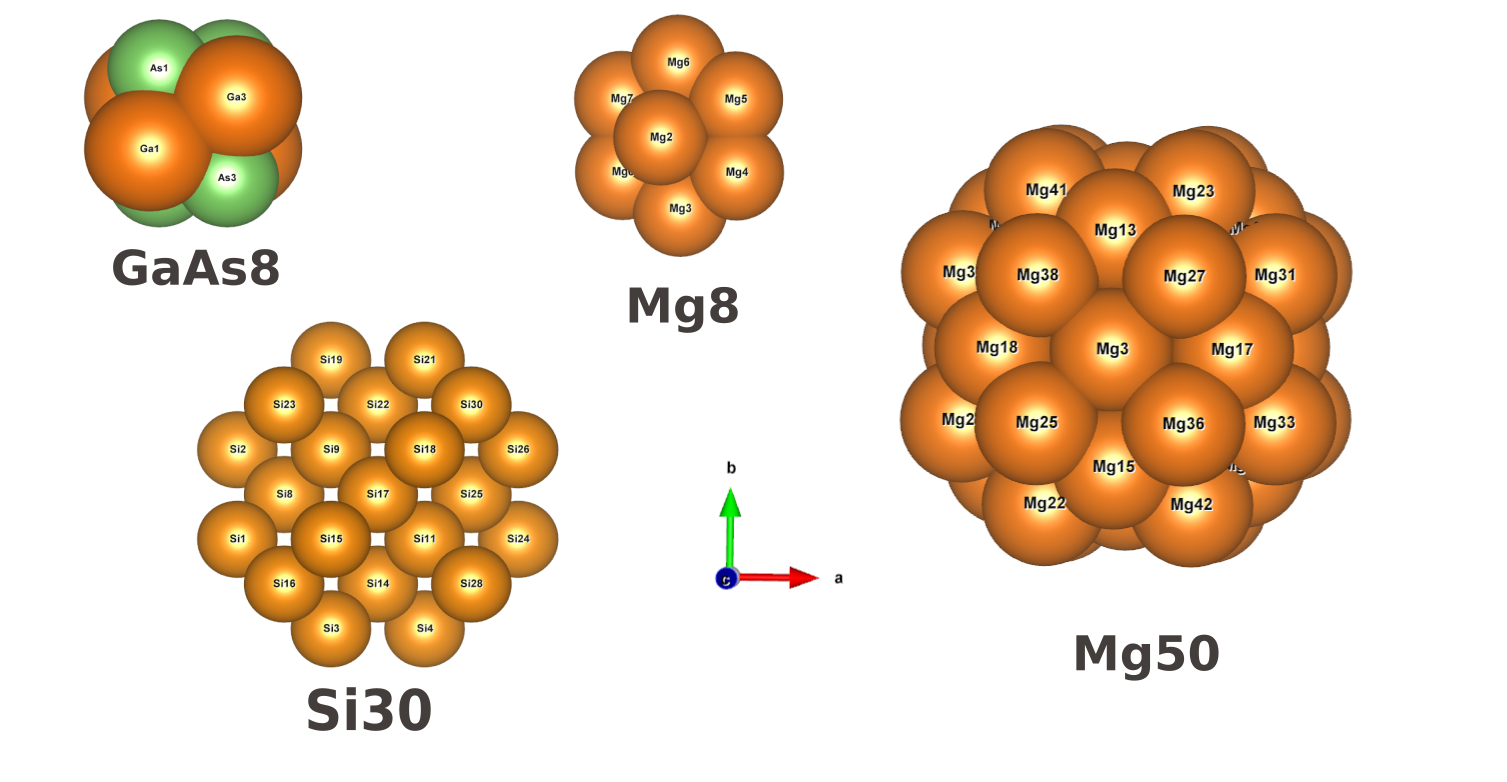}
    \caption{ Cluster structures constructed from CALYPSO code with various system size used in this calculation.}
    \label{fig:cluster_structure}
\end{figure}

As indicated in the previous section, the Pauli constraints result in a range of Eg values. However, to make the method feasible, one needs to fix the value of the $E_g$ parameter. To do so we have considered the benchmark calculations performed for set of (i) metal clusters (Al$_8$, Li$_8$, Li$_{30}$, Mg$_8$, Mg$_{30}$, Mg$_{50}$) and (ii) semiconductor clusters (Si$_8$, Si$_{30}$, Al$_4$P$_4$, Al$_4$Sb$_4$, Ga$_4$As$_4$, Ga$_4$P$_4$, Ga$_4$Sb$_4$, In$_4$As$_4$, In$_4$P$_4$ and In$_4$Sb$_4$). A few of the cluster structures used in the paper are shown in Fig.~\ref{fig:cluster_structure}. Technical details of the calculations are provided in section~\ref{technical_details}.

\subsection{Constrained optimization of $E_g$ and performances}
\label{Eg_min}

\begin{table*}[htbp]
    \centering
    \caption{Relative Median Absolute Relative Errors (RMAREs) of various functionals.}
    \label{tab:RMARE}
    \renewcommand{\arraystretch}{1.3} 
    \setlength{\tabcolsep}{10pt} 
    
    \begin{adjustbox}{max width=\textwidth}
    \begin{tabular}{lccccccc}
        \toprule
        \textbf{Functional} & TFvW & LWT & LMGP ($A=0.2$) & HC ($\lambda=0.01177$, $\beta=0.7143$) & LJGM0 ($E_g=0$) & LJGM ($E_g=0.05$) & LJGM1 ($E_g=0.1$) \\
        \midrule
        RMARE  & 7.13 & 6.67 & 2.28 & 1.71 & 3.09 & 1.42 & 3.75 \\
        \bottomrule
    \end{tabular}
    \end{adjustbox}
\end{table*}

In this section, we first determine the optimal value of the gap parameter $E_g$ employed
in the LJGM kernel. To do so, we systematically optimize $E_g$ by minimizing a global error indicator that simultaneously accounts for errors in total energy and electron
density. Specifically, we employ the relative median absolute relative error (RMARE) as a composite metric to assess the overall performance of the LJGM functional. The RMARE is evaluated relative to the HC~\cite{HC} and LMGP~\cite{LMGP-00} functionals, which are widely regarded as among the most reliable nonlocal KEDFs for localized finite systems, defined as\cite{PGSL_2018_doi:10.1021/acs.jpclett.8b01926}
\begin{equation}
\mathrm{RMARE}
=
\sum_{i}
\frac{\mathrm{MARE}_i}
{\frac{1}{2}\!\left(
\mathrm{MARE}_i^{\mathrm{HC}}
+
\mathrm{MARE}_i^{\mathrm{LMGP}}
\right)},
\label{eq:RMARE}
\end{equation}
where the index $i\in\{\Delta E, D_0\}$ labels the error contributions from the
total energy and the electron density, respectively, defined by,
\begin{eqnarray}
 \Delta E
&=&
\bigl|E^{\mathrm{KS}}-E^{\mathrm{OF}}\bigr|~(\text{for energy})\\
D_0
&=&
\frac{1}{N_e}
\int_{\mathrm{grid}}
\bigl|n^{\mathrm{OF}}(\mathbf r)-n^{\mathrm{KS}}(\mathbf r)\bigr|
\, d^3\mathbf r~(\text{for density})~.
\end{eqnarray}
Thus, all the errors are being calculated with respect to the KS DFT and OFDFT.

Throughout this work, the parameters of the HC functional are fixed at
$\lambda = 0.01177$, $\alpha = 1.952$, and $\beta = 0.7143$, as used in
Ref.~\cite{HC_on_molecoules_10.1063/1.3685604,PGSL_2018_doi:10.1021/acs.jpclett.8b01926,enhanced_vW_WGC} for general use, while the LMGP functional employs $A = 0.2$, consistent with
Ref.~\cite{LMGP-00}.
Lower values of RMARE indicate better overall agreement with KS-DFT in both energy
and density.

\begin{figure}
    \centering
    \includegraphics[width=0.75\linewidth]{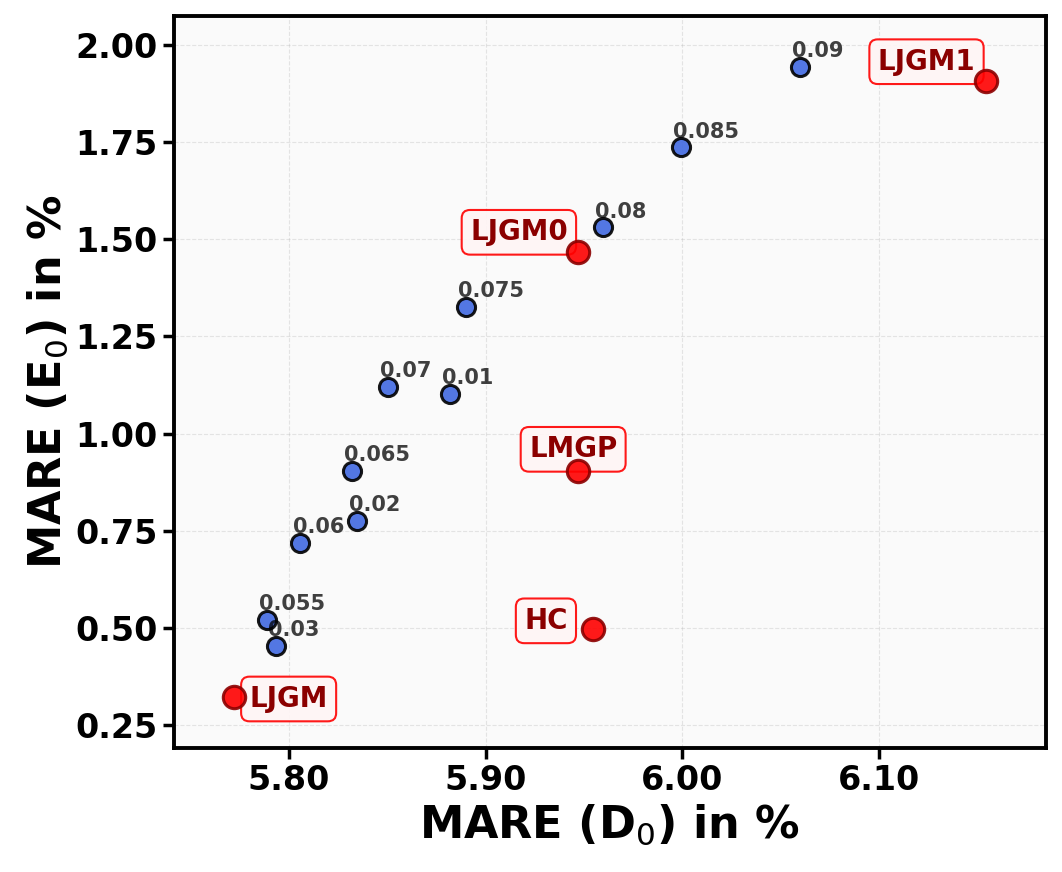}
    \caption{MARE of E$_0$ and D$_0$ for various values of Eg in LJGM compared with HC and LMGP. As mentioned in the text, the labels E$_g$=0.0 eV, 0.05 eV and 0.10 eV denote LJGM0, LJGM and LJGM1 functional respectively. }
    \label{fig:MARE-Eg-opt}
\end{figure}

Based on this optimization procedure, we identify $E_g = 0.05$~eV as the optimal
value for the LJGM kernel, providing the best compromise between accuracy in total
energy and electron density.
Table~\ref{tab:RMARE} summarizes the resulting RMARE values for the different
functionals considered, while Fig.~\ref{fig:MARE-Eg-opt} compares the individual
MARE contributions in $\Delta E$ and $D_0$ across the tested OF-DFT methods.

Fig.~\ref{fig:MARE-Eg-opt} illustrates other cases $E_g = 0.1$~eV
(denoted LJGM1) and $E_g = 0$~eV (denoted LJGM0), the latter corresponding to the
gapless Lindhard kernel.
Eliminating the gap leads to a degradation of the total-energy accuracy by
approximately $2\%$ and a concomitant deterioration in the predicted electron
density. At $E_g = 0.1$~eV, the LJGM functional yields improved density accuracy but at the expense of significantly larger energy errors. These trends confirm that $E_g = 0.05$~eV offers the most balanced and physically meaningful performance for finite systems within the LJGM framework.

\begin{figure}[h!]
    \centering
    \includegraphics[scale=0.5]{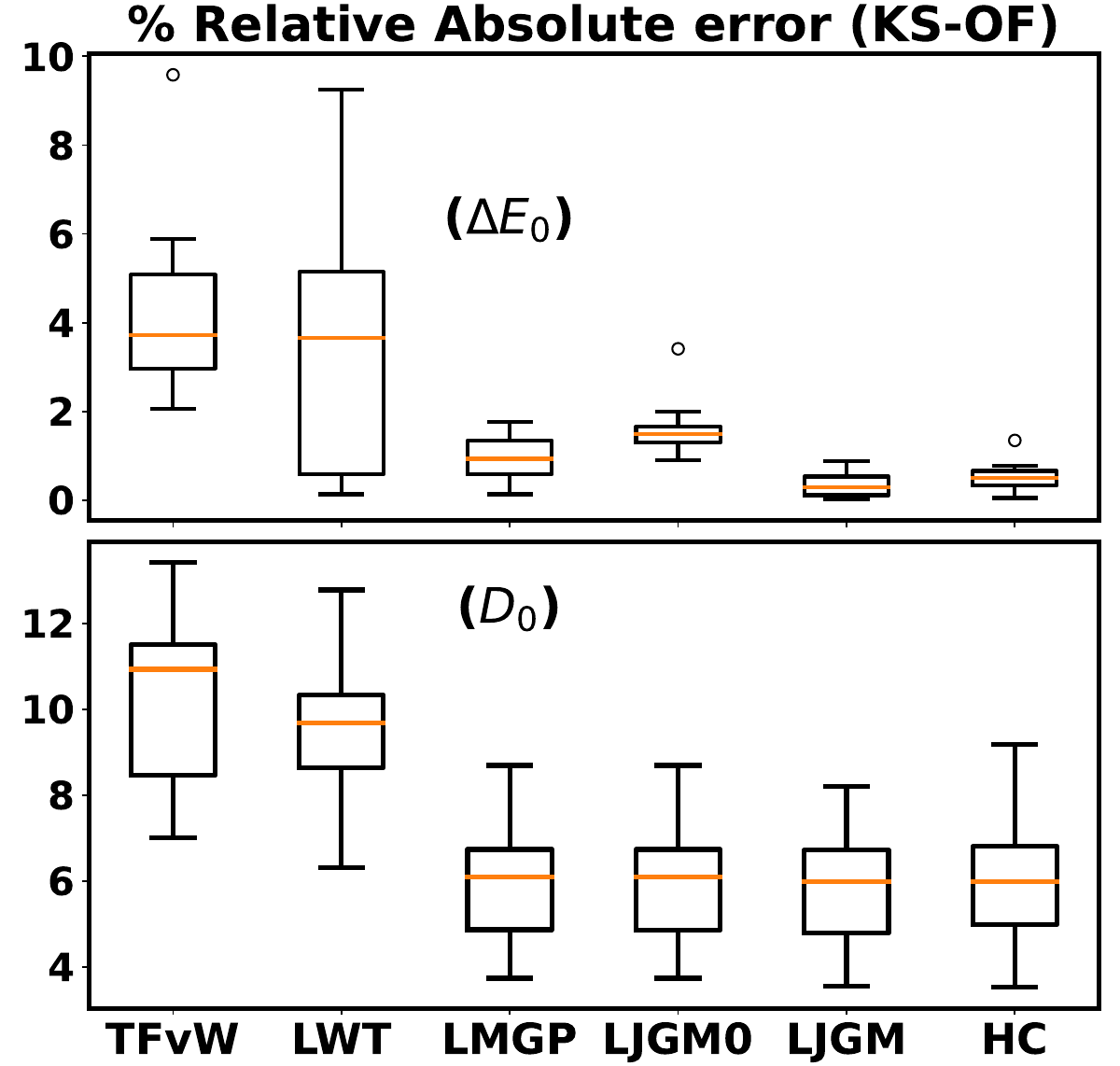}
    \caption{Box plot of absolute relative percentage error (=($|OF-KS|)/|KS|\times 100$) of energy ($E_0$) (in eV) and density error ($D_{0}$) (in a.u.); the band inside the box denotes the median of errors. We have taken 10 semiconductor clusters and 6 metal clusters as mentioned in the section~\ref{sec3:ResultsDiscussion}. The Boxplot used here summarizes the overall distribution of a set of data points in Table S1-S4 of the  supporting information~\cite{support}. The vertical line extends from the minimum to the maximum. }
    \label{fig:Results:Box-RMARE}
\end{figure}

Finally, Fig.~\ref{fig:Results:Box-RMARE} presents a comparative analysis using box plots. These plots illustrate the distribution of errors in total energies and electronic densities across different nonlocal (NL) density-dependent KEDFs. Complementary to this, Table~\ref{tab:RMARE2} also summarizes the Median Absolute Relative Percentage Errors (MAREs) for the complete dataset.

Further, the boxes in Fig.~\ref{fig:Results:Box-RMARE} represent interquartile ranges, with the bottom and top edges corresponding to the first (Q1) and third (Q3) quartiles, respectively. This means that 25\% of the data points fall below Q1, and another 25\% lie above Q3. Outliers, if present, are indicated by circles and are defined as values that deviate more than 1.5 times the interquartile range ($1.5 \times |Q3 - Q1|$) from the box. The vertical whiskers extend from the minimum to the maximum values, excluding outliers.

Using the proposed LJGM functional, the MAREs are 0.32\% for total energy and 5.77 \%  for electron density, representing a clear improvement in density accuracy over HC and LMGP. For comparison, LMGP, among the most accurate orbital-free KEDFs before LJGM, yields MAREs of  0.90\%  (energy) and 5.94\% (density). The next best performers are HC, followed by LJGM0 (LMGP0), as summarized in Table~\ref{tab:RMARE2}. We further note that Moldabekov \emph{et al.}~\cite{Zhandos_Shao_pavanello_Electronic_Structure_2025} reported that OFDFT calculations of Si using the HC functional show measurable discrepancies from KSDFT-derived KE kernels in the long-wavelength regime $q \lesssim 2\pi/a$, even though HC is among the most accurate functionals for semiconductors. The LJGM kernel introduced here addresses this regime by construction: its low-$q$ asymptote $\omega_{T_{NL}}^{\rm LJGM}(q\to 0) \propto \Delta^2(\mathbf{r})/q^2$ [Eq.~\ref{eq:omega_low_q}] directly enforces the correct long-range response, in contrast to HC where this behavior is not built in.

\begin{figure}[htbp]
    \centering
    \includegraphics[scale=0.35]{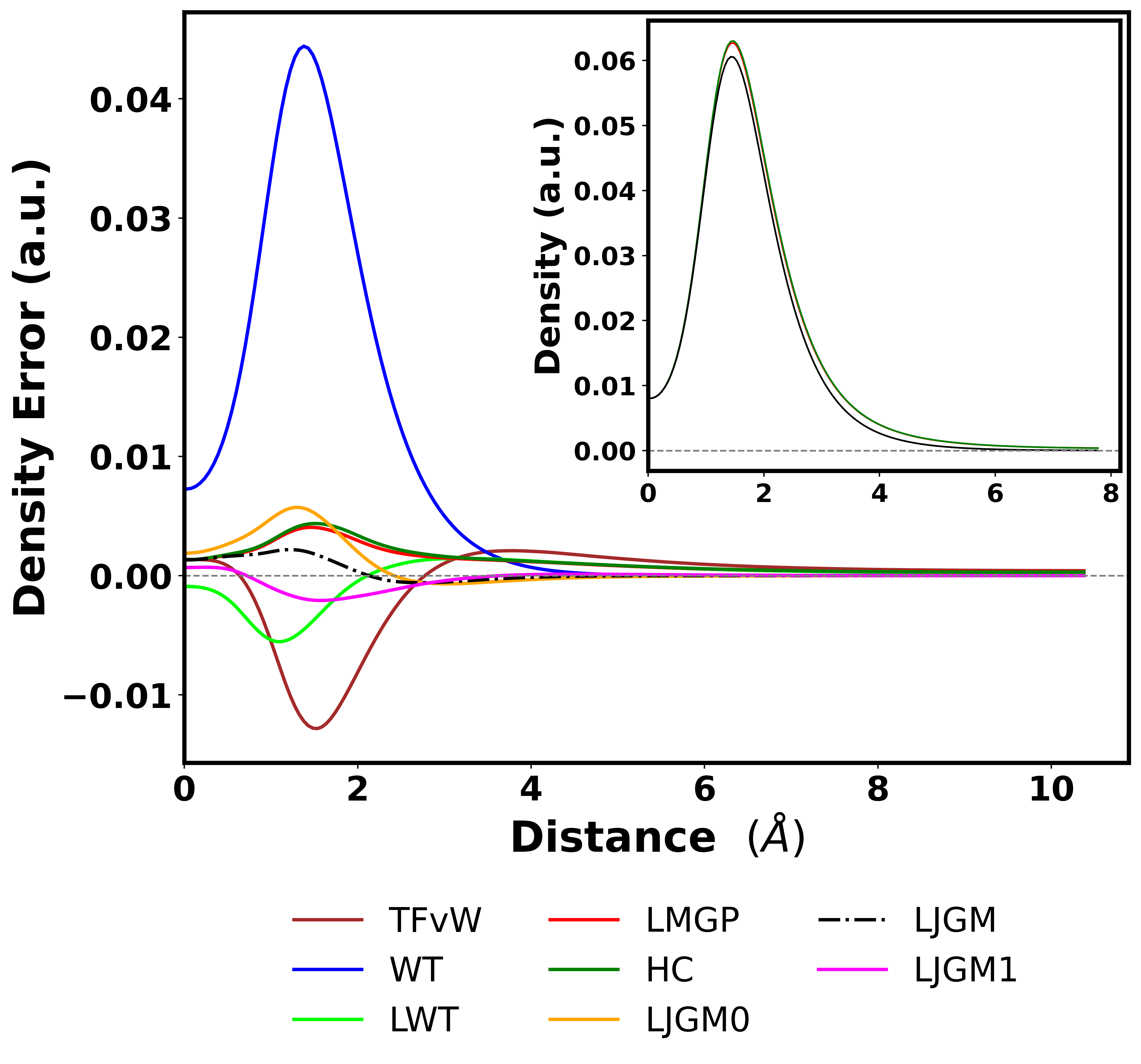}
    \caption{Density difference ($n^{\text{KS}}(\R) - n^{\text{OF}}(\R)$) for Si atom and the inset shows the profile of density. The functionals with density-dependent kernels  are clearly superior to others and behaves similar near nucleus and in tail. The intermediate region where shell structure is dominant, is reproduced best by LJGM with optimal choice of $E_g$. This resonates with our observation in Fig.~\ref{fig:Tp-LJGM-He}. }
    \label{fig:den-Si}
\end{figure}

\begin{figure*}
    \centering
    \includegraphics[scale=0.3]{ 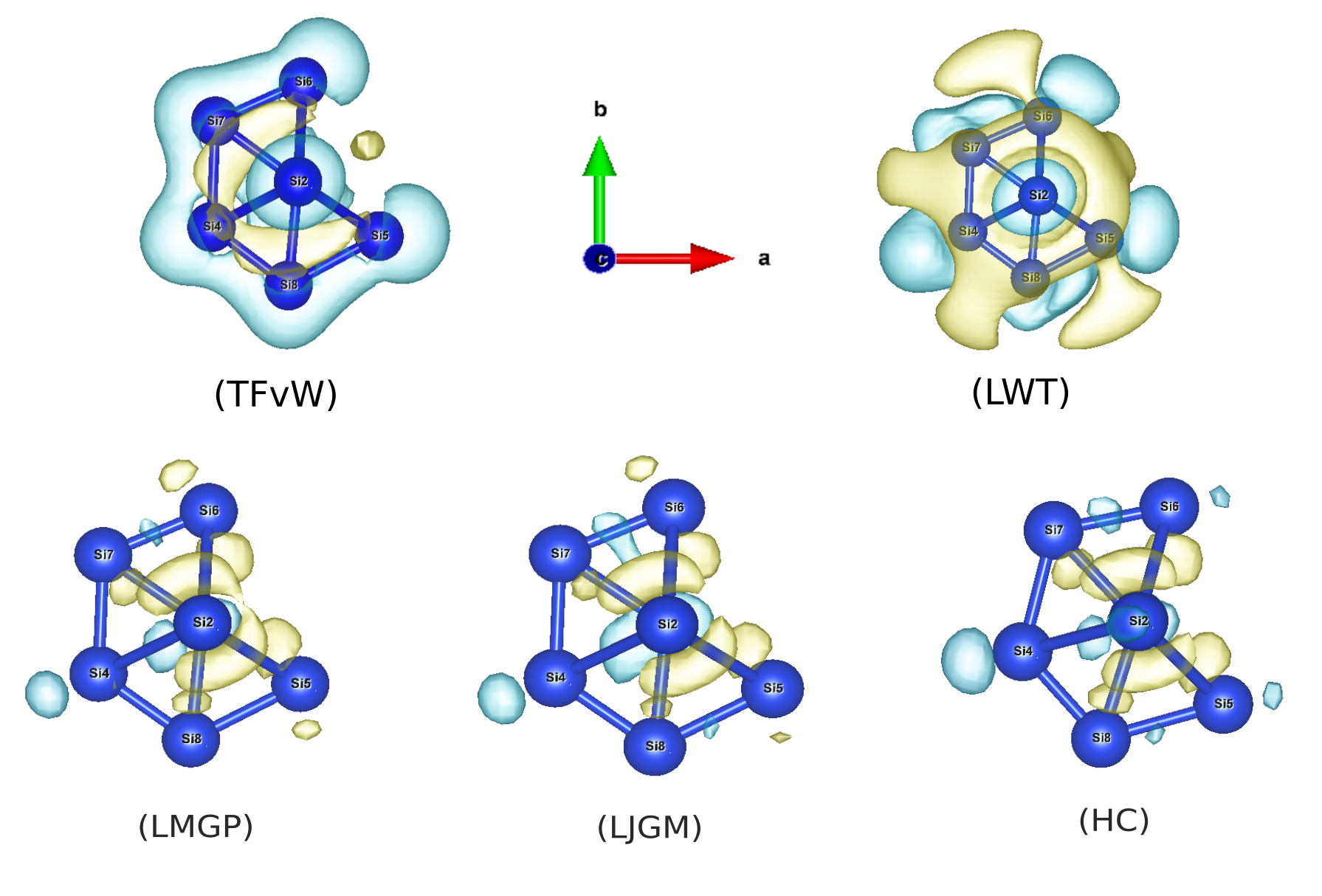}
    \caption{Density difference  ($ n^{KS}- n^{OF}$) of Si$_8$ cluster in 3D in the order: TFvW, LWT, LMGP, LJGM, and HC. We have chosen the top view, displaying how the density improves with increasing functional quality. The yellow shade represents positive values (overestimation) and the blue shade represents negative values (underestimation). We have set the isoscale as: $0.04 (a.u.)\implies 100\%$ yellow saturation and $-0.04 (a.u.) \implies 0\%$ blue saturation.
    }
    \label{fig:3d-dendiff-Si8}
\end{figure*}

To further visualize the accuracy of the density, Fig.~\ref{fig:den-Si} shows the density difference ($n^{KS} - n^{OF}$) for an isolated Si atom using the LJGM functional. The plot demonstrates excellent agreement with the KS result, with the deviation decaying rapidly away from the atomic center. A similar analysis for the Si$_8$ cluster is shown in Fig.~\ref{fig:3d-dendiff-Si8}, which corroborates the findings. For less accurate functionals such as TFvW and LWT, the orbital-free density $n^{OF}$ tends to overestimate the core region and underestimate the interstitial and outer regions. In contrast, the HC, LMGP, and LJGM functionals all show comparable density error profiles for the Si$_8$ cluster, consistent with the trends seen in Fig.~\ref{fig:Results:Box-RMARE}. We have also demonstrated in Fig.S1 and Table.S5 of supporting information~\cite{support} that LJGM and LMGP both have faster rate of convergence and lower computational expense as compared to other KEDFs with density-based kernels.

\begin{table*}[htbp]
    \centering
    \caption{Median Absolute Relative Percentage Errors (MARPEs) of various functionals for semiconductors and metals test set.}
    \label{tab:RMARE2}
    \renewcommand{\arraystretch}{1.3} 
    \setlength{\tabcolsep}{10pt} 
    
    \begin{adjustbox}{max width=\textwidth}
    \begin{tabular}{llcccccccc}
        \toprule
        & & TFvW & LWT & LMGP & LJGM0 & LJGM & LJGM1 & HC \\
        \midrule
        \multirow{2}{*}{\textbf{MARPE}} & \textbf{Energy}   & 3.76  & 3.70  & 0.90  & 1.46  & \textbf{0.32}  &  1.90 & 0.49  \\
                                        & \textbf{Density}  & 10.93 & 9.69  & 5.94  & 5.95  & \textbf{5.77}  &  6.15 & 5.95 \\
        \bottomrule
    \end{tabular}
    \end{adjustbox}
\end{table*}

\begin{figure}[h!]
    \centering
    \includegraphics[scale=0.3]{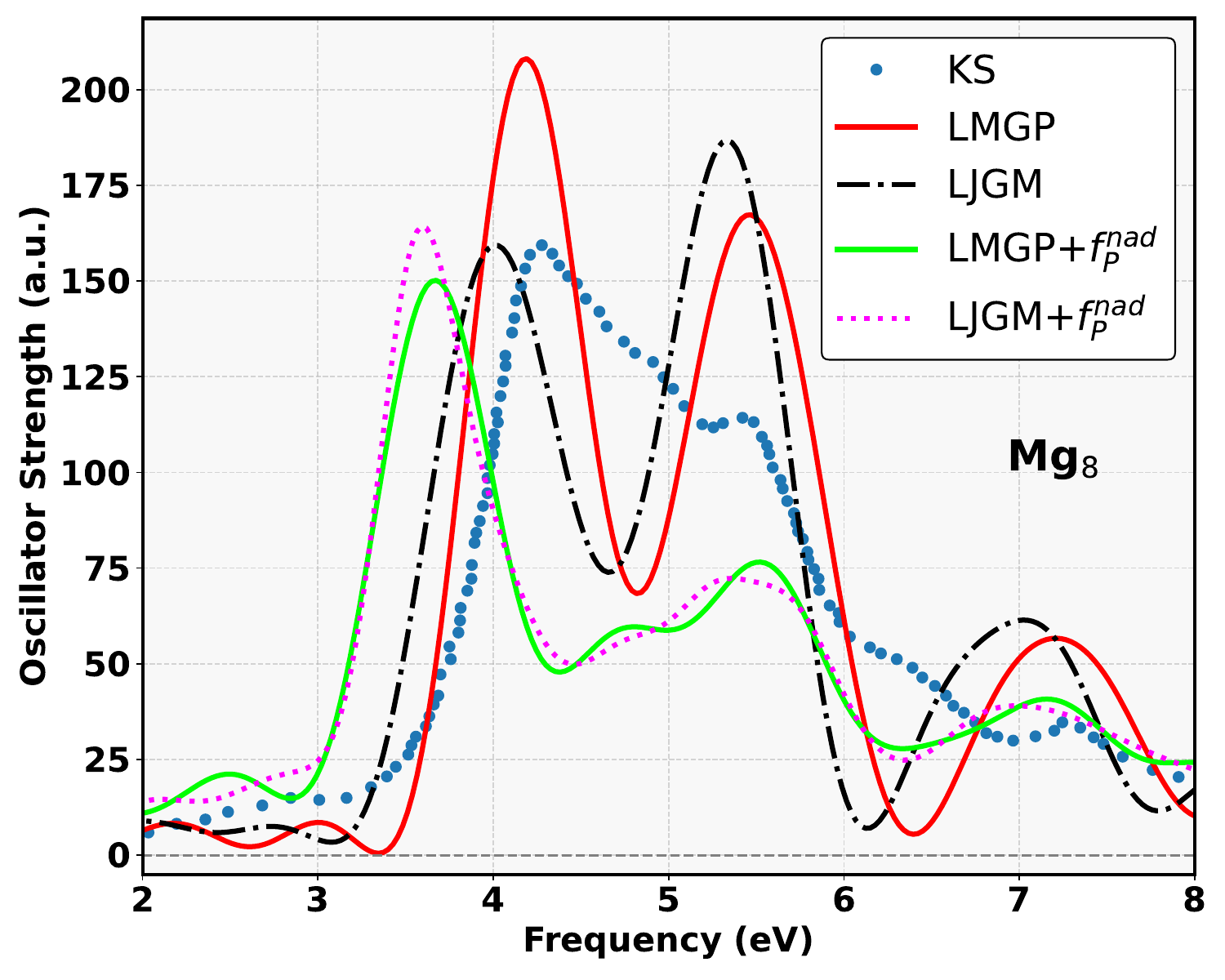}
    \includegraphics[scale=0.3]{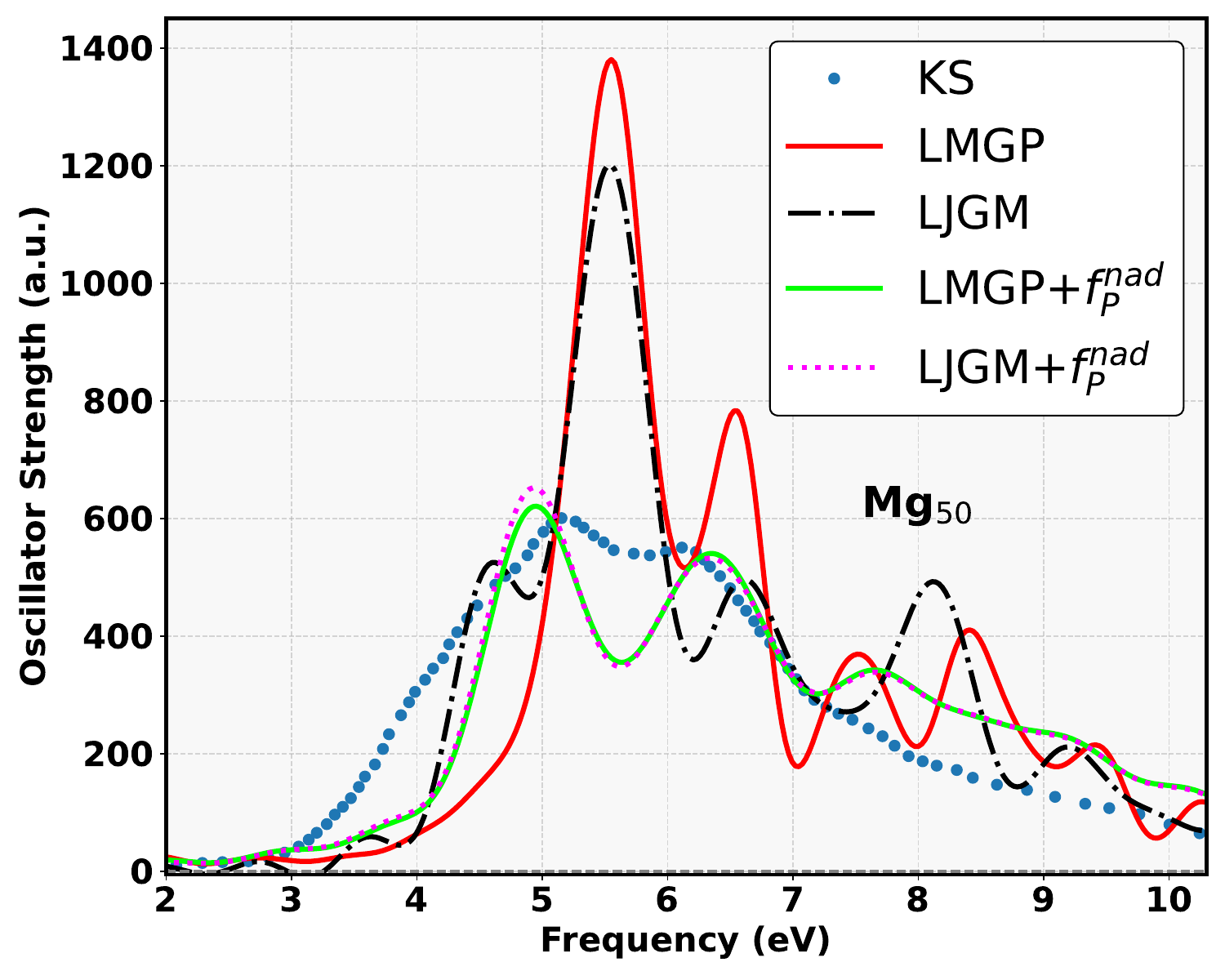}
    \caption{Optical spectrum of Mg$_8$ (left) and Mg$_{50}$ (right) clusters with various approximations. The KS data of Mg$_8$ is taken from ref.~\cite{Pavellano-Kaili-TD-OFDFT_2(2021)_PhysRevB.104.235110}. LMGP+$f^{nad}_P$ and LJGM+$f^{nad}_P$ denote the corresponding spectra with the non-adiabatic Pauli-kernel correction included (eq.~\ref{eq:Sec4-TDOF:JP-Pot}).
    }
    \label{fig:TD-OpticalSpectra}
\end{figure}

\subsection{Optical absorption spectra from Time-dependent orbital-free DFT (TDOFDFT) \label{sec4:Result-TD-OFDFT}}


It is also interesting to show the performance of LJGM together with TD-OFDFT. Before moving to the results, a theoretical background is necessary. Note that unlike KS DFT, the time-dependent OF-DFT (TD-OFDFT) formalism is based on representing a non-interacting bosonic system of $\mathcal{N}$ electrons using a product wavefunction: $\Psi_{B}(\R_{1},\R_{2},\dots, \R_{\mathcal{N}},t)=\prod_{i=1}^{\mathcal{N}} \psi_{B}(\R_{i},t)$, where $\psi_{B}(\R,0)=\sqrt{n^{OF}(\R)}$ and $\Psi_{B}$ denotes the total bosonic wavefunction. In the seminal work by Jiang et. al.~\cite{Pavellano-Kaili-TD-OFDFT_1(2021)_PhysRevB.103.245102}, the foundations of TD-OFDFT were laid by extending adiabatic local density approximation (ALDA) concepts to bosonic systems. It was demonstrated that the essential theorems of TDDFT remain valid, with the Pauli potential $v_{p}(\R,t)$ playing a role analogous to the exchange-correlation (XC) potential in satisfying fundamental constraints.

The TD-OFDFT Hamiltonian takes the form:
\begin{align}
    \hat{\mathcal{H}} = -\frac{1}{2}\nabla^{2} + \underbrace{\frac{\delta T_{s}^{\text{Pauli}}}{\delta n(\R,t)} + v_{s}(\R,t)}_{v_{p} + v_{s} = v_{B}}~,
    \label{eq:Sec4-TDOF:Ham}
\end{align}
where $v_{p}(\R,t) = \delta T_{s}^{\text{Pauli}} / \delta n(\R,t)$ is the adiabatic time-dependent Pauli potential, and $v_{B}$ is the effective bosonic potential. Unlike conventional ALDA, the inclusion of $v_p$ enables TD-OFDFT to account for fermionic effects. Consequently, the accuracy of TD-OFDFT critically depends on the quality of the approximated Pauli potential, as discussed in Sec.~\ref{sec4:Result:pauli_potential}.

The Pauli potential can be decomposed into adiabatic and non-adiabatic parts:
\begin{align}
    v_{p}(\R,t) &= v_{p}^{\text{ad}}(\R,t) + v_{p}^{\text{nad}}(\R,t), \\
    v_{p}^{\text{ad}}(\R,t) &= v_{p}^{\text{GGA}}(\R,t) - v_{p}^{\text{GGA}}(\R,0) + v_{p}^{\text{NL}}(\R,0),
\end{align}
where $v_{p}^{\text{NL}}$ corresponds to the nonlocal Pauli potential derived from orbital-free kinetic energy density functionals (OF-KEDFs). This formulation implies that the accuracy of the non-adiabatic TD-OFDFT largely hinges on the quality of the converged ground-state density and the assumption that the static $v_p^{\text{NL}}$ sufficiently captures the time evolution, an approximation analogous to that of ALDA in TDDFT.

A known limitation inherited from TDDFT is the so-called causality paradox~\cite{TD-OFDFT-causuality-paradox}, arising from the Runge-Gross theorem~\cite{TDDFT-RungeGross-PhysRevLett.52.997}. While various remedies have been proposed~\cite{TDDFT-causuality-resolve-PhysRevLett,TDDFT-Keldysh:1964ud}, the most robust involves including memory effects in the functional~\cite{TD-OFDFT-Vignale-causuality}. It has been demonstrated~\cite{Pavellano-Kaili-TD-OFDFT_1(2021)_PhysRevB.103.245102,Pavellano-Kaili-TD-OFDFT_2(2021)_PhysRevB.104.235110} that such remedies can be incorporated into TD-OFDFT through the concept of a Pauli action integral, replacing the exchange action integral in standard TDDFT.

Since an exact analytic form for $v_p$ is not available, one must resort to approximations. Analogous to the construction of nonlocal KEDFs, $v_p$ in TD-OFDFT can be built using model response functions. Specifically, a gap model kernel such as the JGM has been employed to construct $v_p$ from a nonlocal dynamic response function, thereby correcting the excitonic poles of the non-interacting bosonic system toward those of the HEG~\cite{Pavellano-Kaili-TD-OFDFT_1(2021)_PhysRevB.103.245102}.

In the linear-response regime, the non-adiabatic part of the Pauli potential takes the form:
\begin{align}
    v_{p}^{\text{nad}}(\R,t) = - \mathcal{F}^{-1} \left\{ \vec{q} \cdot \vec{j}(q,t) \frac{\partial f_{p}(q,\omega)}{\partial \omega} \right\},
    \label{eq:Sec4-TDOF:JP-Pot}
\end{align}
where $f_p(q,\omega)$ is the Pauli kernel defined as
\begin{align}
    f_{p}(q,\omega) = \chi_B^{-1}(q,\omega) - \chi^{-1}(q,\omega),
    \label{eq:Sec4-TDOF:fP}
\end{align}
with $\chi_B$ as the bosonic response function and $\chi$ corresponding to either the Lindhard or JGM response. In the JGM approach~\cite{KGAP_dilectric_origin}, the response is defined as
\begin{equation}
\chi_{JGM}(q,\omega) =
\begin{cases}
\chi_{Lind}(q,\omega), & \omega \ge E_g, \\
\displaystyle\frac{k_F}{\pi^2} \frac{1}{F^{JGM}(q,\omega)}, & \omega < E_g,
\end{cases}
\label{eq:Sec4-TDOF:JGM-Real}
\end{equation}
where $F^{JGM}(q,\omega)$ is a model function parametrized by the local Fermi vector $k_F$ and energy gap $E_g$.

Using this, the JGM Pauli kernel takes the form:
\begin{align}
f_{p,JGM}(q,\omega,E_g) &= \chi_B^{-1}(q,\omega) - \chi_{JGM}^{-1}(q,\omega,E_g) \nonumber\\
&= -\frac{3\pi^2}{k_F} \left[ \frac{3}{5} + \frac{q^2}{2k_F^2} + \frac{E_g^2}{k_F^2 q^2} + \frac{12}{175} \frac{k_F^2 q^2}{E_g^2} \right] \nonumber\\
&\quad + \frac{6\pi^2}{k_F^3} \frac{\omega^2}{q^2} \left[1 + \frac{6}{175} \frac{k_F^4 q^4}{E_g^4} \right],
\label{eq:Sec4-TDOF:JGM-fP}
\end{align}
with derivation provided in Appendix~\ref{App3}. The adiabatic and nonadiabatic parts are then:
\begin{align}
f^{\text{nad}}_{p,JGM}(q,\omega,E_g) &= f_{p,JGM}(q,\omega,E_g) - f_{p,JGM}(q,0,E_g) \nonumber \\
&= \frac{6\pi^2}{k_F^3} \frac{\omega^2}{q^2} \left[1 + \frac{6}{175} \frac{k_F^4 q^4}{E_g^4} \right].
\label{eq:Sec4-TDOF:JGM-fP-nad}
\end{align}

Finally, the full nonadiabatic Pauli kernel is given by:
\begin{equation}
f_p^{\text{nad}}(q,\omega) =
\begin{cases}
\text{Eq.~\ref{eq:App1:fpnadLind}}, & \omega \ge E_g, \\
\text{Eq.~\ref{eq:Sec4-TDOF:JGM-fP-nad}}, & \omega < E_g,
\end{cases}
\label{eq:Sec4-TDOF:JGM-Im2}
\end{equation}
and the corresponding Pauli potential is:
\begin{equation}
v_{p}^{\text{nad}}(\R,t) =
\begin{cases}
-\mathcal{F}^{-1} \left\{ \vec{q} \cdot \vec{j}(q,t) \frac{\partial f_{p,Lind}(q,\omega)}{\partial \omega} \right\}, & \omega \ge E_g, \\
-\mathcal{F}^{-1} \left\{ \vec{q} \cdot \vec{j}(q,t) \frac{\partial f_{p,JGM}(q,\omega)}{\partial \omega} \right\}, & \omega < E_g.
\end{cases}
\label{eq:Sec4-TDOF:JGM-Im3}
\end{equation}

This formulation demonstrates how the JGM kernel improves the nonadiabatic Pauli potential in TDOFDFT by modifying the response behavior in the low-frequency regime, which is crucial for describing collective excitations in finite systems.

\subsubsection{Results for TD-OFDFT spectra for clusters}

For TD-OFDFT calculations, the simulation procedure is divided into three main steps: (i) ground-state OFDFT calculation, (ii) real-time (RT) propagation with small time steps, and (iii) incorporation of non-adiabatic corrections via a predictor-corrector scheme during time evolution. The ground-state calculations are performed using the same setup as described in Sec.~\ref{sec3:ResultsDiscussion}. For the real-time propagation, we follow the methodology outlined in Ref.~\cite{DFTpy:Shao_2020}, where a weak external perturbation is applied in the form of a delta-kick with strength $k = 0.1$ a.u. along the $x$-direction.  
All calculations employ the same pseudopotential and XC as mentioned in Sec.\ref{sec3:ResultsDiscussion}. The time interval is set to 0.01 a.u. to ensure both accuracy and stability of the time integration. For both Mg$_8$ and Mg$_{50}$ the convergence is reached below 20,000 iterations. 

The predictor-corrector scheme is applied in real space to propagate the time-dependent dipole moment. The oscillator strength spectrum $\sigma(\omega)$ is obtained from the dipole response via the relation:
\begin{equation}
\sigma(\omega) = -\omega \, \mathrm{Im} \left[ \frac{\mathcal{F}(\delta\mu)}{k} \right],
\end{equation}
where $\delta\mu$ denotes the change in the dipole moment and $\mathcal{F}$ denotes the Fourier transform. To evaluate the performance of the adiabatic Pauli potential, we present the oscillator strength as a function of excitation energy $\omega$ in Fig.~\ref{fig:TD-OpticalSpectra}. The figure shows that the first excitation peak obtained with the LJGM functional aligns more closely with the KS-TDDFT peak, even without incorporating the non-adiabatic correction, highlighting the accuracy of the LJGM ground-state density and its corresponding Pauli potential, as discussed in Sec.~\ref{sec4:Result:pauli_potential}. Specifically, for the Mg$_8$ cluster, LJGM yields a more accurate peak position and spectral shape than LMGP, further validating the improved density quality of the LJGM kernel.

When non-adiabatic corrections are included, both LJGM and LMGP show more pronounced spectral features and closely track each other. This observation aligns with previous findings~\cite{Xuecheng-TDOFDFT_2022,Pavellano-Kaili-TD-OFDFT_1(2021)_PhysRevB.103.245102} that the non-adiabatic contribution often dominates over the adiabatic component in time-dependent orbital-free frameworks. In this work, we include only the leading-order term in the expansion of the non-adiabatic Pauli kernel, i.e., the $\mathcal{O}(\omega)$ term, which renders the resulting Pauli potential frequency-independent. This approximation primarily improves the spectral shape, while the real part of the kernel contributes to the peak shift.

The higher-order correction to the JGM-based Pauli kernel, specifically the $\mathcal{O}(\omega^2)$ term as shown in Eq.~\ref{eq:Sec4-TDOF:JGM-Im2}, has not yet been implemented in the current version of DFTpy. We plan to incorporate this refinement in future work. Notably, even without non-adiabatic Pauli corrections, the LJGM functional accurately predicts the height of the first excitation peak for Mg$_8$ and captures the second excitonic peak of Mg$_{50}$ in a better way, underscoring its reliability and effectiveness. In fact we observe that, with only LJGM kernel, the shift of peak in Mg$_8$ is little red-shifted from KS peak  which is also the case when we add non-diabatic correction Eq.(~\ref{eq:Sec4-TDOF:JGM-Im3}) to   LMGP and LJGM. This hints that the LJGM kernel is inherently capturing the shift of poles in correct direction.

\begin{table}[h]
\centering
\caption{Comparison of the peak position(pos) (eV), oscillator strength ($\sigma(\omega)$) (in a.u) of Mg clusters (units are eV) from Fig.~\ref{fig:TD-OpticalSpectra}. Values in parenthesis are relative error from KS in same unit. }
\begin{tabular}{lcccccccccccc}
\toprule
 & \multicolumn{2}{c}{KS-TDDFT}  &  \multicolumn{2}{c}{LMGP} & \multicolumn{2}{c}{LJGM} & \multicolumn{2}{c}{LMGP+JP} & \multicolumn{2}{c}{LJGM+JP} \\
\cmidrule(r){2-3} \cmidrule(r){4-5} \cmidrule(r){6-7} \cmidrule(r){8-9} \cmidrule(r){10-11} 
System & pos & $\sigma(\omega)$ &  pos & $\sigma(\omega)$ & pos & $\sigma(\omega)$ & pos & $\sigma(\omega)$ & pos & $\sigma(\omega)$ \\
\midrule
Mg$_{8}$ & 4.28 & 158 &  4.16 & 208 & 4.03 & 159 & 3.67 & 150 & 3.59 & 163 \\
  & - & - &  (-0.12) & (50) & (-0.25) & (1) & (-0.61) & (-8) & (-0.69) & (5)
\\
Mg$_{50}$ & 5.10 & 605  & 5.56 & 1378 & 5.512 & 1198 & 4.94 & 621 & 4.96 & 655 \\
 & - & - &  (0.46) & (773) & (0.41) & (593) & (-0.16) & (16) & (-0.14) & (50) \\
\bottomrule
\end{tabular}
\end{table}


\section{Conclusion \label{sec5:Conclusion}}

We have demonstrated the applicability of the JGM model to finite systems such as atomic and molecular clusters, highlighting the importance of an accurate response function, particularly in the low-$q$ limit. We tested the performance of the local JGM-based functional (LJGM) across a broad set of clusters and found LJGM achieves the best overall balance of energy accuracy, density fidelity, and Pauli behavior among tested NL-KEDFs for finite systems. 
Notably, with the LJGM kernel, the correct low$-q$ behavior is recovered from a much more physically motivated ground with no need for any explicit modeling term. And yet LJGM delivers energy and density predictions that closely match those from KS DFT and even better than state-of-the-art OFDFT KEDFs. 
Among all tested density-dependent KEDFs, LJGM exhibited the lowest relative error in cluster total energies, making JGM-model an promising alternative to HEG model.

The LJGM functional also performs well in satisfying exact physical constraints associated with the Pauli potential, providing insight into how effectively the Pauli exclusion principle is captured within an approximate KEDF framework. Importantly, LJGM upholds Pauli positivity in low-density regions, a regime where traditional Lindhard-based functionals typically fail.

To further validate its applicability, we extended the use of LJGM to the calculation of optical spectra. The resulting spectra show significant improvement and align closely with those obtained from KS-TDDFT, especially in the prediction of excitation peak positions and intensities. These results suggest that LJGM is a promising functional for describing finite systems, with potential applications in light-matter interactions, ultrafast processes, and out-of-equilibrium electron dynamics.

\section{Technical Details of Calculations}
\label{technical_details}

All geometries of the most stable cluster-structures are predicted by the CALYPSO code \cite{calypso-code}, which is further used in the Vienna Ab-initio Simulation Package (VASP)~\cite{VASP-1-code,VASP-2-code,VASP-3-code} for geometry optimization. All the OF-DFT calculations have been conducted using the DFTpy software package~\cite{DFTpy:Shao_2020}. The LJGM kernel depends on the local Fermi wavevector $k_F(r)$ and would otherwise require evaluation at every spatial grid point. To retain quasi-linear scaling, we follow the cubic spline interpolation strategy of Mi et al. \cite{LMGP-00} as implemented in DFTpy ~\cite{DFTpy:Shao_2020}. We have used LDA-level bulk-derived local pseudopotentials (BLPS)~\cite{BLPS-HC,DFTpy:Shao_2020} of Huang and Carter~\cite{HC} together with the LDA exchange-correlation~\cite{LDA-XC-PhysRevB.23.5048} potential in all our calculations. A kinetic energy cutoff (ecut) of $1200$ eV and convergence criteria (econv) of $10^{-6}$~eV  is used for all the calculations. We also noted that the convergence rate of LJGM was substantially faster than HC, with convergence criteria being attained in around 100 iterations. A fine grid mesh of $100 \times 100 \times 100$ is considered for the density plot. The KS calculations are performed in Quantum Espresso (QE)~\cite{QuantunEsspresso-code} code  and 408 eV energy cutoff are adopted for well-converged total energies $10^{-7}$~eV in the same simulation cell as the OF-DFT simulations. The KS equations are solved within the LDA (PW92) XC approximation, using the same BLPS-PP of Carter as used in OFDFT calculations generated consistently within the same functional.
The exact Kohn--Sham Pauli kinetic energy density $t_\theta(r) = \tau_s(r) - \tau_{vW}(r)$ and the exact KS Pauli potential $v_{T_\theta}(r)$ shown as reference in Fig.~3 are computed using a modified version of the APE atomic code~\cite{ape_code:OLIVEIRA2008524,LKT-SBTrichey-GGA_PhysRevB.98.041111} in post-SCF orbital-free mode (\texttt{oftype=1}) with the same BLPS pseudopotentials and LDA-PZ exchange-correlation as used in our DFTpy OFDFT calculations. The Pauli potential is reconstructed from the converged KS valence orbitals via the Bartolotti--Acharya construction~\cite{Pauli_pot}.

\section*{Data Availability \label{sec7:data_availability_statement}}
The data supporting the findings of this article are available freely in  supporting information~\cite{support}. 



\section*{Acknowledgments \label{sec6:Acknowlegement}}
The authors thank Prof. Michele Pavanello for the valuable discussions. 
S.\'S. acknowledges the financial support from the National Science Centre, Poland (grant no. 2021/42/E/ST4/00096).

\section*{Supporting Information \label{Supporting}}
The Supporting Information is available free of charge at \cite{support}.
\noindent It contains: per-cluster raw energies for all functionals used to 
construct Fig.~\ref{fig:Results:Box-RMARE} (Tables S1, S2); 
mean-based MARE and RMARE values complementing the median-based metrics 
reported in the main text (Tables S3, S4); convergence behavior of LJGM 
for representative clusters (Figure S1); and computational wall-time 
comparison across functionals and system sizes (Table S5).

The modified version of DFTpy-2.0.0 with the full implementation of LJGM is freely available at~\cite{DFTpy_LJGM_2026}.

\appendix

\section{Pauli positivity condition with JGM-models  \label{App1:PauliPositive} }

In the formulation of OFDFT, the static dielectric function ($\chi$) connects with kinetic energy via the nonlocal kernel $\omega_{T_{NL}}(q)$ of  Eq. \ref{eq:Kernel1}, which is a dimensionless quantity. We wish to explore how this kernel and, consequently, kinetic energy are affected in various density regimes for both Lindhard-based kernels and with the Jellium-with-gap model kernel. In LJGM kernel, $\Delta({\bf{r}})=\frac{2E_{g}}{k_{F}^{2}(\R)}$ 
(with $k_{F}(\R)=(3\pi^{2}n(\R))^{1/3}$) and $\eta=\frac{q}{2k_{F}(\R)}$. Note that we here aim to present a generic proof with jellium with gap model kernel, which shall hold true for both JGM-functional and the present LJGM-functional.

\subsection{ \textbf{Case A: } Low density limits}

Expanding $\omega^{LJGM}_{T_{NL}}$ around  $k_F(\R)\to 0$ we obtain, 

\begin{align} 
    \lim_{k_{F}\to0}\omega_{T_{NL}}(q) &\approx \frac{5}{9\alpha\beta}\Big\{ \frac{3E_{g}^{2}}{k_{F}^{2} q^{2}} + \frac{8}{5}\frac{2E_g^2 - q^4}{4E_g^2 + q^4} 
     -\frac{24}{175} (2 q)^{2}\frac{(24E_{g}^{4} -54E_{g}^{2}q^{4}+q^{8}}{(4E_g^2 + q^4)^{3}}k_{F}^{2} +\mathcal{O}(k_{F}^{3}) \Big\} 
\label{eq:App2:LkFKappa1}
\end{align}
which upon considering $E_g\to 0$ and $\alpha=\beta=5/6$ becomes, 
\begin{align}
    \lim_{\substack{k_F \to 0 \\ E_g \to 0}}\omega_{T_{NL}}(q) &\approx \frac{4}{5} \big\{ -\frac{8}{5} - \frac{96}{175 q^{2}}k_{F}^{2} - \frac{128}{125}\frac{k_{F}^{4}}{q^{4}} \big\} 
    \label{eq:App2:LkFKappa}
\end{align}
One may crosscheck Eq.\ref{eq:App2:LkFKappa} has similar form to Eq.~B5 of Ref.~\cite{KGAP_2018}. 
Now from Eq.~\ref{eq:App2:LkFKappa1}, we consider two cases:

$\mathbf{Case~I: (q\to0)}$
\begin{align}
    \lim_{\substack{k_F \to 0 \\ q \to 0}}\omega_{T_{NL}}(q) &\approx \frac{4}{5} \Big( \frac{3E_g^2}{k_F^2  q^2} + \frac{4}{5} + \frac{36}{175}\frac{k_{F}^{2}}{E_g^2}q^2 + \mathcal{O}(q^6) \Big)
    \label{eq:App2:LkFLqKappa}
\end{align}

$\mathbf{Case~II: (q\to\infty)}$
\begin{align}
    \lim_{\substack{k_F\to0 \\ q\to\infty}}\omega_{T_{NL}}(q) &\approx \frac{4}{5} \Big( \frac{3 E_g^2}{k_F^2 q^2} - \frac{8}{5}-\frac{96}{175} \frac{(q^8 - 54q^4 E_g^2+24E_g^4)}{(4E_g^2+q^4)^3}k_F^2 q^2 + \mathcal{O}(k_F^3) \Big) \approx -1.28 [ 1 + q^{10}(...)~ ]
    \label{eq:App2:LkFHqKappa}
\end{align}
From Eq.~\ref{eq:App2:LkFHqKappa}, 
one can see that this behavior is the same as in the case of Lindhard (see Eq.B1 of Ref.\cite{Shao(2024)-effective-WT-PhysRevB.110.085129}). Interestingly, the constant term being negative and greater than 1 will violate the Pauli positivity condition here: for both Lindhard and Jellium-with-gap model. We emphasize, however, that this is a formal statement about the asymptotic kernel and does not imply that the integrated Pauli potential $v_{\theta}(r)$ becomes negative in real systems. Direct numerical computation of $v_{\theta}(r)$ on our benchmark set (Figs. 2 and 3) shows that $v_{\theta}(r)\ge0$ is preserved throughout the spatial domain for all clusters considered, indicating that the asymptotic violation does not translate into observable pathology in practice.

However in contrast the low-density $q\to\infty$ limit does not have this issue as the leading order and constant term preserves positivity which essentially means the Pauli potential will also be positive.

\subsection{\textbf{Case B: } High density limits}

Similarly one can show that, for high density limit $n(\R)\to\infty$,
\begin{align}
    \lim_{\substack{k_F \to \infty}}\omega^{Lind}_{T_{NL}} = -\frac{2 q^2}{3k_{F}^2}+\frac{q^4}{90 k_{F}^{4}}
\end{align}
which is definitely negative in the low-$q$ limit and might be positive in the high-$q$ limit. Also, in a high-density regime, the vW response should dominate the nonlocal response, and KE should match GE2 in this limit
\begin{align}
    T_{NL}^{Lind} &\approx \int \int n^{\alpha}(r)\mathcal{F^{-1}}[-\frac{8}{9} \Big(\frac{3q}{2k_F} \Big)^{2} + \mathcal{O}\Big(\frac{1}{k_F}\Big)^{4} ]~n^{\beta}(\R') ~dr~d\R' \\
    &= -\frac{8}{9}T_{vW} \\
    \lim_{\substack{n \to \infty}}T_{\theta} & = \lim_{\substack{n \to \infty}}T_{TF} + \lim_{\substack{n \to \infty}}T_{vW} \approx -\frac{8}{9}T_{vW}
\end{align}
Therefore, $T_{\theta}$ is negative in the high-$n$ limit. However the total K.E. $T_{S}$ will be positive, $\lim_{\substack{n \to \infty}}T_{S}=T_{\theta}+T_{vW}=T_{TF}+\frac{1}{9}T_{vW}$ which is nothing but GE2.
Now for JGM-model-based kernels,

\begin{eqnarray}
    \lim_{\substack{k_F \to \infty}}\omega^{JGM}_{T_{NL}} &=& \frac{\pi^2-4}{4} \Big(\frac{E_g}{k_{F}q}\Big)^2 -\frac{2q^2}{3k_F^2} + \frac{\pi}{2}\Big(\frac{E_g}{k_{F}q}\Big)+ \frac{3\pi^2-16}{48} \frac{E_g^2}{k_F^4} + \frac{(64-36\pi^2+3\pi^4)}{48}\Big(\frac{E_g}{k_{F}q}\Big)^4 \nonumber \\  
    &+& \frac{1}{90}\frac{q^4}{ k_{F}^{4}} + \pi\frac{(\pi^2-8)}{8}\Big(\frac{E_g}{k_{F}q}\Big)^3 + \mathcal{O}(k_F^{-6})\nonumber \\
    &=& \frac{\pi}{2}\Big(\frac{E_g}{k_{F}q}\Big) - \frac{2}{3}\frac{q^2}{k_F^2} + \frac{\pi^2-4}{4} \Big(\frac{E_g}{k_{F}q}\Big)^2 + \mathcal{O}(k_F^{-4})
\end{eqnarray}

So from the first line, setting $E_{g}=0$ recovers the result of Lindhard-based kernel, which is expected; And for $E_{g}\neq0$, truncating the series for $\mathcal{O}(k_F^{-4})$, we may note that for low$-q$ limit, the last term dominates and is positive. And in the high-$q$ limit, the second term dominates just like the Lindhard case. Consequently, the leading order in K.E. is $\lim_{n\to\infty}T_{S}\approx T_{TF} + \frac{1}{9}T_{vW}$.

\section{\label{App2:Scaling-constraint} Derivation of scaling constraints in nonlocal KEDFs}
Unlike semilocals, the scaling conditions for NL-KEDFs are not trivial due to their nonlocal forms. The two scaling constraints mentioned in texts are: 
\begin{align}
    T_{\theta}[n_{\lambda}] &= \lambda^2 T_{\theta}[n] \nonumber \\
    v_{T_\theta}[n_\lambda](\mathbf{r}) &= \lambda^2 v_{T_\theta}[n](\lambda \mathbf{r})  \nonumber
\end{align}
The coordinate transformation $\R\to \textbf{z}=\lambda \R$, the density scales as $n(\R)\to n_{\lambda}(\R)=\lambda^3 n(\lambda\R)$ and $k_{F}(\R)\to k_{F_\lambda}=\lambda k_{F}(\lambda\R)$. The expression of Pauli kinetic energy is $T_{\theta}[n]=T_{TF}[n]+T_{NL}[n(\R),n(\R')]$, where $T_{TF}$ is straightforward to show the scaling condition $T_{TF}[n_{\lambda}]=\lambda^2T_{TF}[n]$. Hence we focus on the nonlocal kernel form, we need to know how $\omega$ transforms with coordinate scaling. The Fourier transform operators transform too,
\begin{align}
    \mathcal{F} &=\int d^{3}\R~\exp[-i~q.(\R-\R')] \nonumber\\
    &= \frac{1}{\lambda^3} \int d^3\textbf{z}\exp[-i~p.(\textbf{z}-\textbf{z}')]; ~~~\text{where}~p=\frac{q}{\lambda}
\end{align}
Thus, $\mathcal{F}\longrightarrow  \mathcal{F_{\lambda}}=\lambda^3\mathcal{F}$ and similarly $\mathcal{F}^{-1}\longrightarrow  \mathcal{F_{\lambda}}^{-1}=\frac{1}{\lambda^3} \mathcal{F}^{-1}$.
In the last step the fourier variable transforms as $q\to p=\frac{q}{\lambda}$ this also implies some variable are invariant: $\eta=\frac{q}{2k_{F}}\to\eta_{\lambda}=\frac{p}{2k_{F_{\lambda}}}=\eta$. We borrow the generic form from eq.~\ref{eq:Kernel1},
\begin{align}
    \omega(q)&=\mathcal{F}[\frac{\delta^2 T_{s}}{\delta n(\R)\delta n(\R')}] =\frac{5 ~G_{\text{NL}}(\eta)}{9 \alpha \beta~ n_0^{\alpha + \beta - 5/3}}\nonumber \\
    &=\lambda^{3(\alpha+\beta)-5} \frac{5 ~G_{\text{NL}}(\eta_{\lambda})}{9 \alpha \beta~ n_0^{\alpha + \beta - 5/3}}; ~~~ \text{since}~~\eta_{\lambda}=\eta \nonumber \\
    \omega(q) &= \omega_{\lambda}(p); ~~~ \text{putting } \alpha=\beta=5/6
    \label{eq:scaled_omega}
\end{align}
In the last step we gave used the value of $\alpha+\beta=5/3$ as used in for LWT, LMGM and LJGM. Now this translates to Fourier space as, $\omega(k_{F}|\R-\R'|) = \mathcal{F}^{-1}[\omega(q)]=\lambda^3 \mathcal{F_{\lambda}}^{-1}[\omega(p)]=\lambda^3 \omega_{\lambda}(\textbf{z}-\textbf{z}')$.
\begin{align}
    T_{NL}[n(\R),n(\R')] &= \int d^3\R d^3\R' n^{\alpha}(\R)\omega(k_{F}|\R-\R'|)n^{\beta}(\R') \nonumber\\
    &= \int \frac{1}{\lambda^{6}} d^3\textbf{z}d^3\textbf{z}' \lambda^{3(\alpha+\beta)} n_{\lambda}^{\alpha}~~\lambda^{3} \omega_{\lambda}(k_{F_\lambda}|\lambda\R-\lambda\R'|) n_{\lambda}^{\beta} \nonumber \\
    &= \lambda^{2} \int d^3\textbf{z}d^3\textbf{z}'~ n_{\lambda}^{\alpha} \omega_{\lambda}(k_{F_\lambda}|\textbf{z}-\textbf{z}'|) n_{\lambda}^{\beta} \nonumber \\
    T_{NL}[n(\R),n(\R')]&= \lambda^2 T_{NL}[n(\textbf{z}),n(\textbf{z}')] 
    \label{eq:scaled_T_proved}
\end{align}
In the last step, we have used $\alpha+\beta=5/3$ and we get condition 4 for scaling: $T_{\theta}[n_{\lambda}]=\lambda^2 T_{\theta}[n]$.

Now to prove condition 5 we must start from generic form of nonlocal potential (eq.~\ref{eq:jgm_potential}),
\begin{align}
    v_{T_{\text{NL}}}(\mathbf{r}) &= \frac{1}{n(\mathbf{r})^{1/6}}  
    \mathcal{F}^{-1} \Big[ \mathcal{F} \left[ n(\mathbf{r})^{5/6} \right] \omega(\eta) \Big] \\
    &= \frac{1}{\lambda^{1/2}~ n_{\lambda}(\textbf{z})} \mathcal{F_{\lambda}}^{-1} \Big[ \mathcal{F_{\lambda}} \left[ \lambda^{5/2}~n(\mathbf{z})^{5/6} \right] \omega_{\lambda}(\eta_\lambda) \Big] \\
    v_{T_{NL}}(\R) &= \lambda^{2} v_{T_{NL}}(\textbf{z})
\end{align}
 In the last step we have used $\omega(q)=\omega_{\lambda}(p)$. One may note that satisfying scaling constraints for nonlocal KEDFs entirely depend on how $\omega$ is scaled. 

\section{\label{App3} Details of the non-adiabatic kernels for TD-OFDFT}

This appendix shows the details of the non-adiabatic kernels based on Lindhard and jellium with gap model kernels for TD-OFDFT. The generalized form of Pauli-kernel in real-space is defined as:
\begin{align}
    f_{P}(\R,\R',t,t') = \Big[ \frac{\delta v_{P}(\R,t)}{\delta n(\R',t')}\Big]_{n=n_{0}}
\end{align}
In Fourier space, it corresponds to 
\begin{eqnarray}
   f_{P}(q,\omega) &= \mathcal{F}^{-1}\Big[ \int d(t-t')~ e^{i\omega(t-t')} f_{P}(\R,\R',t-t') \Big]    
\end{eqnarray}
Using a Dyson-like equation~\cite{Giuliani_Vignale_2005} in Fourier space, one can also define $f_{P}(q,\omega)$ in terms of the Bosonic system (B) and the non-interacting system (S)~\cite{Pavellano-Kaili-TD-OFDFT_2(2021)_PhysRevB.104.235110} 
\begin{equation}
f_{P}(q,\omega) = \chi_{B}^{-1}(q,\omega) - \chi_{S}^{-1}(q,\omega)~.
\label{bb1}
\end{equation}
Following ref.~\cite{Pavellano-Kaili-TD-OFDFT_2(2021)_PhysRevB.104.235110} for Bosonic system
\begin{align}
    \chi_{B}^{-1}(q,\omega) &= \mathcal{F} \big[\int d(t-t')~e^{i\omega(t-t')}~\chi_{B}(\R,\R',t-t') \Big] = \frac{3\pi^{2}}{k_{F}^{3}} \Big( \frac{1}{\omega-q^{2}/2+i \gamma} - \frac{1}{\omega + q^{2}/2 + i \gamma} \Big)^{-1}
    \label{eq:App1:chiB1}
\end{align}

\subsection{Lindhard}
As the analytic expression of $\chi_{S}(q,\omega)$ is not known hence one has to approximate with the frequency-dependent Lindhard function~\cite{Giuliani_Vignale_2005} in terms of variables $\tilde{\eta} (=\frac{q}{2k_{F}}), 
 \tilde{\omega} (=\frac{\omega}{qk_{F}}), \tilde{\gamma} (=\frac{\gamma}{qk_{F}}))$, which gives the following approximations~\cite{Giuliani_Vignale_2005}, 
\begin{eqnarray}
    \chi_{Lind}^{-1}(q,\omega) &=& \frac{\pi^{2}q}{k_{F}^{3}} \Big[ -\tilde{\eta} + \frac{1}{4} \big(1-(-\tilde{\eta}+i \tilde{\gamma}+\tilde{\omega} )^{2} \big)\log(\frac{1-\tilde{\eta}+i \tilde{\gamma}+\tilde{\omega} }{-1-\tilde{\eta}+i \tilde{\gamma}+\tilde{\omega} }) \nonumber\\
    &-& \frac{1}{4}\big(1-(\tilde{\eta}+i \tilde{\gamma}+\tilde{\omega} )^{2} \big)
    \log(\frac{1+\tilde{\eta}+i \tilde{\gamma}+\tilde{\omega} }{-1+\tilde{\eta}+i \tilde{\gamma}+\tilde{\omega} }) \Big]^{-1}
    \label{eq:App1:chiLind}
\end{eqnarray}
And in the same prescription, $\chi_{B}^{-1}$ of Eq.~\ref{eq:App1:chiB1} simplifies to
\begin{align}
    \chi_{B}^{-1} &= \frac{6\pi^{2}}{k_{F}}\tilde{\eta} \Big[ \frac{1}{\tilde{\omega}-\tilde{\eta} + i \tilde{\gamma}} - \frac{1}{\tilde{\omega}+\tilde{\eta} + i \tilde{\gamma}} \Big]~.
    \label{aa1}
\end{align}

The non-adiabatic kernel is now defined as,
\begin{equation}
f_{P}^{nad}(q,\omega)=f_{P}(q,\omega)-f_{P}(q,\omega=0)~,
\end{equation}
which after substituting 
Eq.~\ref{aa1} and Eq.~\ref{eq:App1:chiLind} into Eq.~\ref{bb1} and taking (i) $q\to 0$ and $\omega\to 0$ and after (ii) putting $\gamma\to 0$ results,
\begin{eqnarray}
f_{P, Lind}^{nad}(q,\omega)&=&\frac{\pi^{2}}{k_{F}} \Big[\frac{i\pi}{6}(\frac{3}{qk_{F}}+\frac{2q}{4k_{F}^{3}})\omega+ \frac{(16-\pi^{2})}{4q^{2}k_{F}^{2}}\omega^{2}+ \frac{16-3\pi^{2}}{48 k_{F}^{4}}\omega^{2} \Big]     \label{eq:App1:fpnadLind}
\end{eqnarray}
This is the resultant equation of the non-adiabatic kernel in the Lindhard framework. Also, the reader is suggested to go through Ref.~\cite{Pavellano-Kaili-TD-OFDFT_1(2021)_PhysRevB.103.245102} for details.

\subsection{Jellium with gap model}

In the following we consider the same procedure to establish the non-adiabatic kernel with Jellium with gap model.  
Following Ref.~\cite{KGAP_dilectric_origin} the frequency-dependent dielectric function, $\epsilon(q,\omega)=\epsilon_1 + i\epsilon_2$ obeys,
\begin{align}
    \epsilon_1(q,\omega) =
    \begin{cases}
        \epsilon_1^{Lind}(q,\omega), &  \omega \ge \lambda\omega_F, \\
        \epsilon_1^{JGM} , &  \omega < \lambda\omega_F
    \end{cases}
\end{align}
and 
\begin{align}
    \epsilon_2(q,\omega) =
    \begin{cases}
        \epsilon_2^{Lind}(q,\omega), &  \omega \ge \lambda\omega_F, \\
        0 , &  \omega < \lambda\omega_F~,
    \end{cases}
\end{align}
where (considering atomic units) $\omega_F=E_{F}\approx\frac{k_F^2}{2}$, and $\lambda=\frac{E_g}{\omega_F}$ is a dimensionless quantity. If $\lambda=0$, then the JGM dielectric function becomes the Lindhard function. Again $\epsilon_1^{JGM}$ is defined as,
%
\begin{eqnarray}
    \epsilon_1(q\omega) &=& 1 + \frac{2}{k_{F}\pi}\Big[\frac{1}{Q^2}-\frac{\Delta}{2Q^3}\big\{\tan^{-1}(\frac{2Q+Q^2}{\Delta} + \frac{2Q-Q^2}{\Delta} ) \big\} \nonumber\\
    &+& \{\frac{\Delta^2}{8Q^2}+\frac{1}{2Q^3}-\frac{1}{8Q}\}\ln{ \big[ \frac{\Delta^2+(2Q+Q^2)^2}{\Delta^2+(2Q-Q^2)} \big] } \Big]
\end{eqnarray}
%
where $\Delta^2=\lambda^2-\frac{\omega^2}{\omega_F^2}\approx\frac{4 (Eg^2-\omega^2)}{k_F^4}>0$, and $Q=\frac{q}{k_F}$.

Now the dielectric and response are related via: $\epsilon(q,\omega)=1-v(q)\chi(q,\omega)$, and substituting $Q=2\tilde{\eta}$ we obtain
%
\begin{eqnarray}
\epsilon_1(q,\omega)-1 &=& \frac{2}{\pi k_{F}} \frac{1}{8\tilde{\eta}^2} \Big[ \frac{1}{2}-\frac{\Delta([\tan^{-1}(\frac{4\tilde{\eta}+4\tilde{\eta}^{2}}{\Delta})+ \Delta)[\tan^{-1}(\frac{4\tilde{\eta}-4\tilde{\eta}^{2}}{\Delta})]}{8\tilde{\eta}}\nonumber\\
&+& (\frac{\Delta^{2}}{128\tilde{\eta}^{3}} + \frac{1}{8\tilde{\eta}} - \frac{\tilde{\eta}}{8} ) \ln\Big[ \frac{\Delta^{2}+(4\tilde{\eta}+4\tilde{\eta}^{2})^{2}}{\Delta^{2}+(4\tilde{\eta}-4\tilde{\eta}^{2})^{2}} \Big]\nonumber\\
&=&\frac{1}{\pi k_{F}} \frac{1}{4\tilde{\eta}^2} [F^{-1}_{JGM}]
\end{eqnarray}

%
Now from the definition $-v(q)\chi(q,\omega)= \frac{k_F}{\pi}\frac{1}{q^2}[F^{-1}_{JGM}]$, and we know that $v(q)\propto\frac{1}{q^2}$, we obtain $\chi_{JGM}^{-1}(q,\omega)=\frac{\pi}{k_F}F_{JGM}$ (using put $\tilde{\eta}=\frac{q}{2k_F}$ ). 

Now expanding the JGM response around $\tilde{\eta}\to 0$ and $\tilde{\omega}\to 0$, and setting $\gamma=0,~ \lambda=\frac{2E_g}{k_F^2}$ we obtain

\begin{align}
    \chi^{-1}_{JGM}(\tilde{\eta},\tilde{\omega},E_{g}) &= \frac{3 \pi^2}{k_F} \left[\frac{3}{5} + \frac{E_g^2}{4 \eta_1^2 k_F^4} + \eta_1^2 - \frac{48 k_F^4 \eta_1^2}{175 E_g^2} - \omega_1^2 \left(\frac{q^2}{4 \eta_1^2 k_F^2} + \frac{48 k_F^6 q^2 \eta_1^2}{175 E_g^4}\right)
\right]
\end{align}

\begin{align}
    \lim_{\tilde{\eta}\to0}\chi^{-1}_{JGM}(\tilde{\eta},\tilde{\omega}=0,E_{g}) &= \frac{3\pi^{2}}{k_{F}} \Big[ \frac{3}{5} + \frac{\lambda^{2}}{16\tilde{\eta}^{2}} + \tilde{\eta}^{2}-\frac{192}{175}\frac{\tilde{\eta}^{2}}{\lambda^{2}} \Big]
\end{align}
Following the same procedure, one can derive $f_{P,JGM}^{nad}(q,\omega,E_{g})=(\chi_{B}^{-1}(q,\omega) - \chi_{B}^{-1}(q,0))-(\chi_{JGM}^{-1}(q,\omega)-\chi_{JGM}^{-1}(q,0))$

\begin{align}
    f_{P,JGM}^{\text{nad}}(\tilde{\eta}, \tilde{\omega}, E_{g}) &= \frac{3 \pi^2 \tilde{\omega}^2}{k_F} 
    + \frac{3 \pi^2 \tilde{\omega}^2 q^2}{4 \tilde{\eta}^2 k_F^3} 
    + \frac{144 \pi^2 \tilde{\omega}^2 \tilde{\eta}^2 k_F^5 q^2}{175 E_g^4} \\
    &= \frac{6\pi^{2}}{k_{F}^{3}} \frac{\tilde{\omega}^{2}}{q^{2}} \left[ 1 + \frac{6}{175} \frac{k_F^4}{E_{g}^{4}} q^4 \right]~.
    \label{eq:App1:fpnadJGM}
\end{align}
Eq.~\ref{eq:App1:fpnadLind} along with Eq.~\ref{eq:App1:fpnadJGM} and with some algebra we arrive Eq.~\ref{eq:Sec4-TDOF:JGM-Im2} of the main text. As one can see, the real part of JGM-adiabatic does not have any linear contribution of $\omega$. Note that we here present a generic proof with jellium with gap model kernel, which shall hold true for both JGM-functional and the present LJGM-functional.

\clearpage
\bibliography{reference}

\end{document}